\documentclass[traditabstract]{aa} 
\usepackage{graphicx}
\usepackage{url}
\usepackage{lscape}
\usepackage{longtable}
\usepackage{multirow}
\usepackage{txfonts}

\newcommand{\hi}{H\textsc{i}}   

\begin{document}
   \title{Cold gas properties of the \textit{Herschel} Reference Survey. III. Molecular gas stripping in cluster galaxies}


  \author{A. Boselli\inst{1}
  	  ,
	  L. Cortese\inst{2,3}
	  ,
	  M. Boquien\inst{1}
	  ,
	  S. Boissier\inst{1}
	  ,
	  B. Catinella\inst{2,4}
	  ,
	  G. Gavazzi\inst{5}
	  ,
	  C. Lagos\inst{3}
	  ,
	  A. Saintonge\inst{6}
         }
	
	\authorrunning{Boselli et al.}
	\titlerunning{Molecular gas stripping in cluster galaxies} 
	 
\institute{	
	Laboratoire d'Astrophysique de Marseille - LAM, Universit\'e d'Aix-Marseille \& CNRS, UMR7326, 38 rue F. Joliot-Curie, 13388 Marseille Cedex 13, France 
        \email{Alessandro.Boselli@lam.fr; mboquien@ast.cam.ac.uk; Samuel.Boissier@lam.fr}
	\and
	Centre for Astrophysics \& Supercomputing, Swinburne University of Technology, Mail H30, PO Box 218, Hawthorn, VIC 3122, Australia
        \email{lcortese@swin.edu.au; bcatinella@swin.edu.au}
        \and
	European Southern Observatory, Karl-Schwarzschild Str. 2, D-85748 Garching bei Muenchen, Germany
        \email{clagos@eso.org}
	\and
	Max-Planck-Institut f\"{u}r Astrophysik, D-85741 Garching, Germany
	\and
	Universita degli Studi di Milano-Bicocca, Piazza della Scienza 3, 20126, Milano, Italy 
	\email{giuseppe.gavazzi@mib.infn.it}
	\and
	Max-Planck-Institut fur Extraterrestrische Physik, D-85741 Garching, Germany
	\email{amelie@mpe.mpg.de}
        }

   \date{}

 
  \abstract
  {The \textit{Herschel} Reference Survey is a complete volume-limited, K-band-selected sample of nearby objects
  including Virgo cluster and isolated objects. Using a recent compilation of H{\sc i} and CO data for this sample 
  we study the effects of the cluster environment on the molecular gas content of spiral galaxies. With the
  subsample of unperturbed field galaxies, we first identify the stellar mass as the scaling variable that traces the total molecular gas mass of galaxies better.
  We show that, on average, H{\sc i}-deficient galaxies are significantly offset (4 $\sigma$) from the $M(H_2)$ vs. $M_{star}$ relation for H{\sc i}-normal galaxies.
  We use the $M(H_2)$ vs. $M_{star}$ scaling relation to define the H$_2$-deficiency parameter as the difference, on logarithmic scale, between the expected and observed molecular gas mass
  for a galaxy of given stellar mass. The H$_2$-deficiency parameter shows a weak and scattered relation with the H{\sc i}-deficiency parameter, 
  here taken as a proxy for galaxy interactions with the surrounding cluster environment. We also show that, as for the atomic gas, the extent of the molecular disc 
  decreases with increasing H{\sc i}-deficiency. All together, these results show 
  that cluster galaxies have, on average, a lower molecular gas content than similar objects in the field.
  Our analysis indicates that ram pressure stripping is the physical process responsible for this molecular gas deficiency.
  The slope of the $H_2-def$ vs. $\mathrm{\hi}-def$ relation is less than unity, while the $D(\mathrm{\hi})/D(i)$ vs. $\mathrm{\hi}-def$
  relation is steeper than the $D(CO)/D(i)$ vs. $\mathrm{\hi}-def$ relation, thereby indicating that the molecular gas is removed less efficiently than the atomic gas. This 
  result can be understood if the atomic gas is distributed on a relatively flat disc that is more extended than the stellar disc. It is thus less anchored
  to the gravitational potential well of the galaxy than the molecular gas phase, which is distributed on an exponential disc with a scalelength $r_{CO}$ $\simeq$ $0.2 r_{24.5}(g)$.
  There is a clear trend between the $NUV-i$ colour index, which is a proxy for the specific star formation activity, and the H$_2$-deficiency parameter, which suggests 
  that molecular gas removal quenches the activity of star formation. This causes galaxies migrate from the blue cloud to the green valley and, eventually, to the red sequence.
  The total gas-consumption timescale of gas deficient cluster galaxies is comparable to that of isolated, unperturbed systems. The total gas depletion timescale determined by considering
  the recycled fraction is $\tau_{gas,R}$ $\simeq$ 3.0-3.3 Gyr, which is significantly larger than the 
  typical timescale for total gas removal in a ram pressure stripping process, indicated by recent hydrodynamical simulations to be $\tau_{RP}$ $\simeq$ 1.5 Gyr. 
  The comparison of these timescales suggests that ram pressure, rather than a simple stop of the infall of pristine gas from the halo, will be the dominant process 
  driving the future evolution of these cluster galaxies.

  }
   {}
   {}
   {}
   {}
   {}

	       
   \maketitle
%

\section{Introduction}

The molecular gas phase plays a fundamental role in the process of star formation in galaxies. The atomic gas, which is dominant in normal, field late-type galaxies, collapses within molecular clouds to form new stars.
This simple matter cycle can be significantly perturbed in dense environments (e.g. Boselli 2011). The atomic gas phase, which in normal, unperturbed galaxies extends up to $\sim$ 1.8 times the 
stellar disc (Cayatte et al. 1994; Bigiel \& Blitz 2012), can be easily removed during the interaction of galaxies with the surrounding environment (Boselli \& Gavazzi 2006). 
Since the seminal work of Haynes \& Giovanelli (1984), there has indeed been strong observational evidence that late-type cluster galaxies have less atomic gas than similar objects in the field 
(e.g. Cayatte et al. 1990; Solanes et al. 2001; Gavazzi et al. 2005). Tidal interactions with nearby companions
or with the cluster gravitational potential well (galaxy harassment, Moore et al. 1998), the dynamical interaction between the hot intracluster medium (ICM) with the cold interstellar medium (ISM)
of galaxies moving at high velocity ($\sim$ 1000 km s$^{-1}$) within clusters
(ram pressure stripping, Gunn \& Gott 1972; thermal evaporation, Cowie \& Songaila 1977) can remove the gaseous component. 

The lack of gas naturally quenches star formation, thereby transforming active systems in
quiescent objects (e.g. Boselli \& Gavazzi 2006). The timescales relative to this transforming process, and the structural, physical and kinematical properties of the perturbed galaxies drastically change according to the perturbing process.
Gravitational interactions indistinctly perturb all galaxy components (dark matter, stars, gas, dust, etc.) thereby producing strong signatures in the morphological and kinematical 
structure of the perturbed objects. To be efficient in fully transforming the whole late-type galaxy population falling into rich clusters, however, 
they require very long timescales since multiple encounters are necessary. Indeed, given the high-velocity dispersion within clusters ($\sim$ 1000 km s$^{-1}$), the perturbative phase
during a fly-by encounter is too short to induce strong effects (Boselli \& Gavazzi 2006). The dynamical interactions with the hot, dense ICM and, in particular, the 
ram pressure stripping phenomenon, has been identified as the dominant process perturbing late-type galaxies in nearby rich clusters (e.g. Vollmer et al. 2001; Boselli \& Gavazzi 2006; Tonnesen et
al. 2007). Observations, models and simulations all indicate that the atomic gas phase can be easily perturbed 
on relatively short timescales ($\sim$ 100-200 Myr; Vollmer et al. 2004; Roediger et al. 2005; Boselli et al. 2006; 2008a, 2008b; Crowl \& Kenney 2008). 
The effects on the molecular gas phase are still under study. Modelling the effects of ram pressure on a multiphase gas disc is still challenging
since it requires taking several physical effects into account, such as heating and cooling of the ISM, self-gravity, star formation,
stellar feedback, and magnetic field. Given the fractal distribution of the gaseous component within the ISM, models must also simultaneously predict
the gas distribution on different scales, from giant molecular clouds to the tails whose size can exceed the size of galaxies (Roediger 2009). Recent attempts to simulate 
the effects of ram pressure on a multiphase ISM on different scales have been carried out by Tonnesen \& Bryan (2009). These authors have shown that the molecular gas phase, 
even in high-density regions, can be removed during the interaction. Whereas detailed multifrequency observations of representative objects
in nearby clusters begin to show cases of ongoing molecular gas ram pressure stripping events (Vollmer et al. 2008a, 2009, 2012a; Sivanandam et al. 2010), 
strong and convincing statistical evidence is still lacking.
Since star formation occurs mainly inside molecular clouds, the perturbation of the molecular gas phase in any kind of interaction can have a strong impact on the evolution of the stellar populations
of galaxies. It is thus paramount to understand whether the molecular hydrogen is perturbed or not and removed in high-density environments.\\

The first studies of the molecular gas content of cluster galaxies and the search for any possible evidence of molecular gas stripping 
come from Kenney \& Young (1989) for the Virgo cluster and from Casoli et al. (1991) for the Coma cluster.
Kenney \& Young (1988; 1989) discovered low-luminosity Virgo cluster spirals characterised by weak CO emission lines, which they interpreted as
the first proof of a low molecular hydrogen content in this cluster. Casoli et al. (1991), on the other hand, observed 
normal molecular gas content in a few spiral galaxies in the Coma cluster and deduced that the cluster environment is not able to
remove the molecular phase, which is strongly anchored to the deep gravitational potential well of galaxies. These results were later questioned by Boselli et al. (1997)
as due to strong biases. Boselli et al. (1997) have shown how the low CO emission of the low-luminosity Virgo cluster galaxies 
discovered by Kenney \& Young (1988) was not evidence of the lack of H$_2$, but rather an effect related to the decrease in the conversion factor $X_{CO}$
in low-luminosity, metal-poor objects with respect to massive, metal-rich galaxies. Indeed, in these
low-metallicity systems the interstellar radiation field is able to dissociate the molecular gas in the diffuse medium, where $X_{CO}$ is low, 
thus increasing the mean $X_{CO}$ factor of these galaxies with respect to massive objects (Boselli et al. 1995, 1997, 2002). The 
result of Casoli et al. (1991) has also been criticised because the observed Coma cluster galaxies were selected in the far-infrared. Given the tight
correlation between the far-infrared and the CO emission, this selection criterion might have excluded molecular gas poor objects from the observed sample (Boselli
et al. 1997, 2002). Other CO surveys of nearby clusters have been published in Lavezzi \& Dickey (1998) and in Scott et al. (2013).
Using a collection of CO data taken from the literature for a large sample of nearby galaxies, Jablonka et al. (2013) have recently shown that, on average, cluster galaxies have a lower CO luminosity per unit 
far-infrared luminosity than field objects. They interpreted this result as an evidence of molecular gas stripping in high-density environments.
This result, however, is based on a heterogeneous sample of isolated and cluster galaxies often selected according to different criteria and
thus might be biased by subtile and complex selection effects (e.g. Boselli et al. 1997; 2013c). 
The lack of cluster and field samples selected according to similar criteria prevented previous works from deriving 
any firm conclusion on this topic. In recent years, several efforts have been undertaken to construct a sample of isolated, unperturbed galaxies with molecular gas data that are
ideally defined to be used as references 
for any statistical study. Sauty et al. (2003) made systematic CO observations of galaxies in the catalogue of isolated objects constructed by Karachentseva (1973). Observations of these 
isolated objects were later extended by Lisenfeld et al. (2011). However, these new datasets have not yet been used to quantify the molecular gas content of nearby cluster galaxies.

In the preparation of the \textit{Herschel} mission (Pilbratt et al. 2010) we began a \textit{SPIRE} guaranteed time key project aimed at studying the properties of the ISM
of a K-band-selected, volume-limited sample of 322 nearby galaxies. This project, named the \textit{Herschel} Reference Survey (HRS), has been extensively described in Boselli et al. (2010). 
To provide the community with the largest possible set of multifrequency data necessary for any kind of statistical analysis, we have collected data at all wavelengths, from UV (Boselli et
al. 2011; Cortese et al. 2012a) to radio centimetric, including H$\alpha$ narrow-band imaging (Boselli et al. in prep.), integrated optical spectroscopy (Boselli et al. 2013a), PACS (Cortese et al. 2014) 
and SPIRE photometry (Ciesla et al. 2012). For such purpose, we have recently obtained new CO data for a large fraction of the late-type galaxies of the HRS. We combined these data 
with those available in the literature and produced a homogenised catalogue of $^{12}$CO(1-0) data for 225 galaxies of the HRS sample. We also collected from
the literature and homogenised H{\sc i} data for 315 out of the 322 HRS galaxies (256/260 for spirals; Boselli et al. 2013b, Paper I). 
The quality of the data at hand, the accurate estimate of the total CO flux emission
and of their uncertainties, and the presence of a well defined, complete sample including cluster and isolated galaxies selected according to similar criteria, 
all allow us to extend and revisit the critical topic of the effects of the environment on the molecular 
gas properties of cluster galaxies on the best statistically significant sample
with CO data available in the literature. The study of the effects of the environment on the atomic gas content of cluster galaxies has been presented in Cortese et al. (2011), the one on
the dust mass in Cortese et al. (2010, 2012). In an accompanying paper (Boselli et al. 2013c; paper II of this series), we study the molecular and total gas scaling relations of the HRS.

\section{The sample}

The analysis presented in this work is based on the spiral galaxies of the \textit{Herschel} Reference Survey, 
a K-band-selected (K $\leq$ 12 mag), volume-limited (15 $\leq$ $D$ $\leq$ 25 Mpc) sample of nearby objects
spanning a wide range in morphological type (from Sa to Sd-Im-BCD) and stellar mass (10$^9$ $\lesssim$ $M_{star}$ $\lesssim$ 3 10$^{11}$ M$_{\odot}$
for those galaxies of the sample with available molecular gas data). The K-band selection was chosen as proxy for stellar mass (Gavazzi et al. 1996).
The sample, which is extensively described in Boselli et al. (2010), is perfectly suited for environmental studies since it includes 
galaxies in high-density regions such as those located in the core of the Virgo cluster and fairly isolated objects in the field. The local galaxy density
of the sample, defined as in Tully (1988) as the number of galaxies brighter than $M_B$ = -16 per Mpc$^3$,
ranges from $\simeq$ 4 gal Mpc$^{-3}$ in the core of the Virgo cluster to $\simeq$ 0.2 gal Mpc$^{-3}$ in the field (Tully 1988). The cluster
environment is also characterised by a diffuse, hot intergalactic medium ($T$ $\simeq$ 10$^7$ K) whose density reaches $\simeq$ 2 10$^{-3}$ atoms cm$^{-3}$
(B\"ohringer, private comm.) and a velocity dispersion of the order of 1000 km s$^{-1}$ (Boselli \& Gavazzi 2006).\\
The analysis done in this work is based on the subsample of spiral galaxies (type$\geq$ Sa) with available CO data (168 objects). As extensively described in the next section,
we use the H{\sc i}-deficiency parameter to quantify the degree of perturbation induced by the cluster environment on the atomic gas component of our sample.

Sixty \% of the analysed galaxies have $\mathrm{\hi}-def$ $\leq$ 0.4 and can thus be considered as unperturbed objects. Indeed, out of these 101 galaxies, 93 are 
located outside Virgo cluster A, the X-ray dominated region surrounding  M87, and B, close to M49, where the perturbations are known to be dominant. The remaining 40 \% 
have a significant lower atomic gas content than objects of similar size and morphological type in the field. These are galaxies preferentially 
located within Virgo cluster A and B (33/67), in the outskirts of Virgo, 
or in the other substructures of the cluster (W, W1, M, East and North clouds, Southern extension) (19/67), where the perturbations are known to be milder (Gavazzi et al. 1999),
with a few objects in close pairs or multiple systems (9/67). 
Seventeen out of the 168 spirals have molecular gas masses estimated from accurate CO mapping done by Kuno et al. (2007) (see Table \ref{TabKuno}). The analysis presented in paper I
has shown that the uncertainty on the total CO flux of these galaxies is $\simeq$ 12 \%, thus significantly smaller than the typical uncertainty on the CO flux of the rest of the sample
($\gtrsim$ 44 \%) determined by applying aperture corrections. Since the present analysis is devoted to the study of second order effects (is the dispersion in the scaling relations 
controlled by the environment?), the use of high-quality data is of paramount importance. We will thus consider in the following analysis the subsample of HRS galaxies observed by Kuno et al. (2007) as a
high-quality sample, and see whether the relations traced by the whole HRS on a large statistical basis are also shared by this subsample of high-quality data. Table \ref{TabKuno} shows that 
the subsample of HRS galaxies observed by Kuno is well suited for this purpose since it includes eight unperturbed objects ($\mathrm{\hi}-def$ $\leq$ 0.4)
and nine gas-poor galaxies, seven of which located within Virgo A. 
The analysis presented in this work has also shown a significant lack of CO data for H{\sc i}-deficient low-mass objects. This lack of data might introduce
systematic effects in the comparison of the statistical properties in the molecular gas content of cluster and field objects. 
Whenever necessary, we thus consider separately the whole sample of galaxies and the subsample of massive objects ($M_{star}$ $>$ 10$^{10}$ M$_{\odot}$) complete in CO data
to test the robustness of our conclusions.

\section{The data}

The data used for the following analysis have been collected from the literature or from our own observations and homogenised to provide complete catalogues
useful for any kind of statistical analysis. Molecular hydrogen masses have been derived from $^{12}$CO(1-0) data available for 168 out of the 260 spiral
galaxies of the sample (143 detections). CO data of 59 out of these 168 objects come from our own observations done at the 12 metre Kitt Peak radiotelescope,
presented in paper I. CO fluxes are converted into molecular gas masses assuming either a constant, Galactic conversion factor of 
$X_{CO}$ = 2.3 10$^{20}$ cm$^{-2}$/(K km s$^{-1}$) (Strong et al. 1988), or the H-band luminosity-dependent conversion factor of Boselli et al. (2002),
log $X_{CO}$ = -0.38 $\times$ log $L_H$ ($L_{H\odot}$)+ 24.23 (cm$^{-2}$/(K km s$^{-1}$).

Complete CO mapping, available for the brightest 37 objects (22\%) and single beam observations 
have been corrected for aperture effects and homogenised as described in Boselli et al. (2013b; paper I). The estimated error on the total, extrapolated CO flux 
ranges from $\sim$ 12\% ~ for the integrated values given in Kuno et al. (2007) (17 galaxies) to $\gtrsim$ 44\% in those objects where the total flux has been
determined from single beam observations reaching $\sim$ 120\% ~ when the surface of the galaxy covered by the beam of the telescope is $<$ 10\%
~ of the total optical surface. Despite the accurate aperture correction, a possible systematic bias can affect the total molecular gas determiation
of the perturbed galaxies.
Indeed, if this extrapolation is very accurate in isolated, unperturbed
galaxies such as those used to define the H$_2$-deficiency parameter (Boselli et al. 2013b), this might not be the case in perturbed cluster objects. The molecular gas distribution of these cluster 
galaxies might be truncated because of their interaction with the cluster environment. Furthermore, in these objects the scalelength of the CO emitting disc 
is not necessarily extended as in unperturbed systems, as is indeed the case for the atomic component. Variations of the molecular radial profile of H{\sc i}-deficient galaxies has 
been indeed observed by Fumagalli et al. (2009). More recently, Davies et al. (2013) have shown that also the molecular gas disc of the few CO detected early-type galaxies 
within the Virgo cluster is less extended than that of similar objects in the field.
The H{\sc i} gas profile is significantly truncated and the mean gas column density reduced in the most deficient cluster galaxies 
as indicated by interferometric observations (Cayatte et al.
1994). NGC 4569, an anemic, gas deficient galaxy ($\mathrm{\hi}-def$ =
1.05) close to the core of the cluster, has a reduced molecular gas column density and a profile truncated at the same level than the H{\sc i} and 
the star forming disc (Boselli et al. 2006). 

The extrapolated CO flux of H{\sc i}-deficient galaxies with single beam observations measured using the prescription 
given in paper I might thus be overestimated, and as a consequence the H$_2$-deficiency parameter underestimated. To quantify this effect, we have extrapolated the observed CO radial profile of NGC 4569 
of Kuno et al. (2007) and compared the integrated CO flux to that determined by limiting the aperture up to the truncation radius. The difference between the extrapolated and the aperture limited flux is 
less than a few percent. We have also tried to correct for aperture effects single beam observations of the HRS galaxies integrating the CO exponential radial profile 
to the isophotal FUV radii given in Cortese et al. (2012a) rather than to infinity.
Indeed at these short wavelengths sensitive to the youngest stellar populations isophotal diameters follow the truncation of the atomic gas disc (Cortese et al. 2012a). Again, 
the difference in the CO integrated profiles using $g$-band (as done in paper I) or FUV isophotal radii is less than a few percent. This test, however, can only be taken as indicative since we
are forced to assume that the scalelength of the CO disc is unchanged between unperturbed and perturbed objects: $r_{CO} = 0.2 r_{24.5}(g)$. 
These tests suggest that any possible systematic bias in the determination of the total CO
emission of perturbed galaxies with single beam observations should be minor. This evidence is explained by the fact that in normal galaxies the CO radial profile is very steep. 
To have a significant impact on the determination of the total CO emission
using the prescription given in paper I, the truncation of the molecular gas disc must occur in the inner region. Multifrequency observations combined with
stripping models of interacting galaxies indicate that this can happens only in dwarf systems (Boselli et al. 2008a).\\

Atomic gas data are available for 256/260 of the spiral galaxies of the sample, and for 100\% of those objects with available CO data. 
The atomic gas data are used to estimate the
degree of perturbation induced by the cluster environment on late-type galaxies. This is done through the determination of the H{\sc i}-deficiency parameter $\mathrm{\hi}-def$,
defined as the difference in logarithmic scale between the expected and the observed
H{\sc i} mass of a galaxy of given angular size and morphological type (Haynes \& Giovanelli 1984). The H{\sc i}-deficiency for all the HRS galaxies 
is determined using the recent calibration of Boselli \& Gavazzi (2009). Atomic and molecular gas data are also used to derive the total cold gas content of the target
galaxies. This is done accounting for 30\% of helium.\\

Optical radii, determined at the 24.5 mag arcsec$^{-2}$ $g$-band isophote, and stellar masses are taken from Cortese et al. (2012a). The latter have been determined
through $i$-band luminosities with the $g-i$ colour-dependent stellar mass-to-light ratio relation from Zibetti et al. (2009), and assuming a Chabrier (2003) initial
mass function. \\

Star formation rates are determined using UV \textit{GALEX} data (from Boselli et al. 2011, Cortese et al. 2012a), corrected for dust attenuation using 22 $\mu$m 
\textit{WISE} data (from Ciesla et al. 2014) following the prescription of Hao et al. (2011). This calibration has been originally defined for
the \textit{Spitzer} MIPS 24 $\mu$m band, but given the similarity of the flux densities in the two adjacent bands (\textit{WISE} 22 $\mu$m and MIPS 24 $\mu$m), 
it can be adopted without any correction. UV corrected luminosities are transformed into star formation rates
(in M$_{\odot}$ yr$^{-1}$) using the standard calibrations of Kennicutt (1998), as described in Boselli et al. (2009). These star formation rates are consistent with those 
determined using narrow band H$\alpha$ imaging data corrected for [NII] contamination and dust attenuation with the Balmer decrement using integrated spectroscopy 
(Boselli et al. 2013a): the mean ratio for the HRS galaxies is log$\frac{SFR(H\alpha)}{SFR(UV)}$=0.04$\pm$0.28. \\

Molecular and total gas masses, combined with star formation rates, are used to estimate the total and molecular gas depletion times, or equivalently total and
molecular star formation efficiencies (Young et al. 1996; Boselli et al. 2001; 2002). The definition and the determination of all these variables is extensively
described in paper II. \\

\section{The molecular gas content of cluster galaxies}

\subsection{Calibration of the H$_2$-deficiency parameter}

The definition of a gas-deficiency parameter was first introduced by Haynes \& Giovanelli (1984). The idea beyond this definition is that 
of quantifying the amount of gas removed from galaxies during their interaction with the surrounding medium (Boselli \& Gavazzi 2006). 
Using a large sample of isolated galaxies, Haynes \& Giovanelli (1984) determined standard scaling relations between the total H{\sc i} content and different characteristic 
variables representative of the size of galaxies.
Among these, they identified the relation between the total H{\sc i} mass and the optical diameters as the one with the smallest scatter (0.3 dex).
This relation can be used to estimate the total mass of H{\sc i} gas that a galaxy belonging to different environments of a given size should have, 
and compare it to the observed value. The difference between the expected and the observed value, in logarithmic scale, is the H{\sc i}-deficiency parameter.
The large sample of isolated galaxies in their hand allowed them to define different calibrations for galaxies of different morphological type.\\

Consistent with what done in our previous works on the molecular gas properties of cluster galaxies (Boselli et al. 1997, 2002),
we use the present set of molecular gas data to define and calibrate a recipe for measuring the molecular gas deficiency of the HRS spiral galaxies.
We first need to identify the unperturbed objects of the sample to be taken as reference for the determination of the calibrating scaling relations. The atomic
gas component generally extends out to $\sim$ 1.8 times the optical disc and is thus poorly anchored to the gravitational potential well. It is thus the galaxy component most
easily perturbed in any kind of interaction (Boselli \& Gavazzi 2006). For this reason we identify the unperturbed objects of the sample using the H{\sc i}-deficiency parameter ($\mathrm{\hi}-def$
$\leq$ 0.4). This threshold in the H{\sc i}-deficiency parameter is chosen because is very close to the typical dispersion in the distribution of $\mathrm{\hi}-def$ observed in
isolated galaxies ($\simeq$ 0.3; Haynes \& Giovanelli 1984). 

\begin{figure*}
   \centering
   \includegraphics[width=14cm]{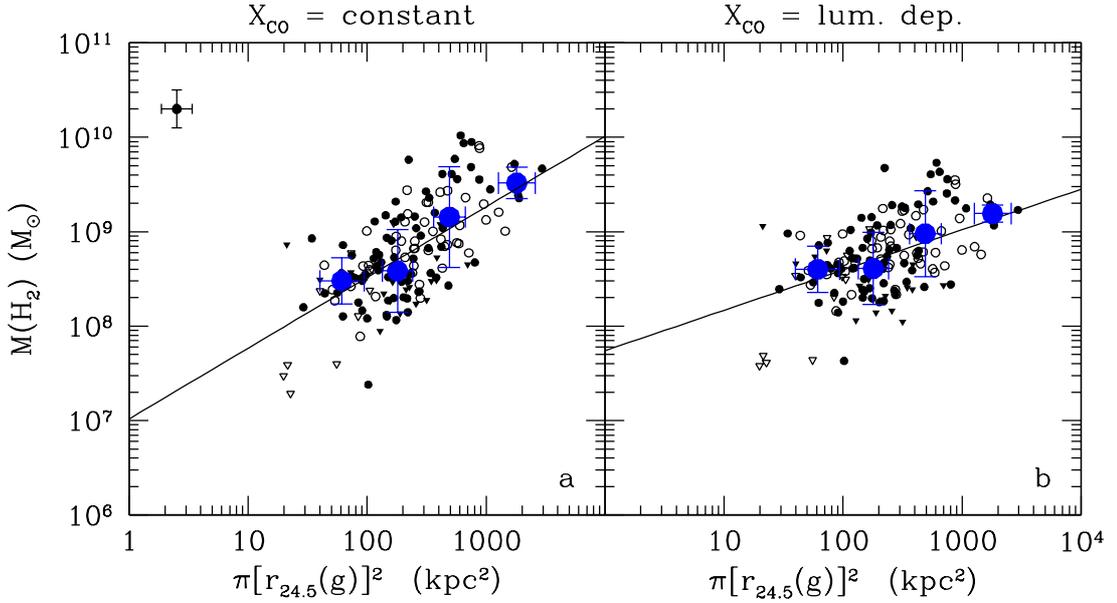}
   \caption{The relationship between molecular gas and the optical surface of spiral galaxies
   when the molecular gas mass is determined assuming a constant $X_{CO}$ factor (a, left) or the luminosity-dependent $X_{CO}$ factor 
   given in Boselli et al. (2002) (b, right). 
   Circles indicate detected galaxies, triangles upper
   limits. Filled symbols indicate 
   objects with a normal H{\sc i} gas content ($\mathrm{\hi}-def$ $\leq$ 0.4), empty symbols H{\sc i}-deficient objects ($\mathrm{\hi}-def$ $>$ 0.4). 
   The typical error bars on the data is given in panel a. The big blue filled dots  
   indicate the mean values and the standard deviations for the whole sample with $\mathrm{\hi}-def$ $\leq$ 0.4.
   The black solid line indicates the linear fit determined using all gas-rich galaxies ($\mathrm{\hi}-def$ $\leq$ 0.4).
   Mean values and best fits have been done including non-detected CO galaxies at their upper limit.}
   \label{calraggi}%
   \end{figure*}

\begin{figure*}
   \centering
   \includegraphics[width=14cm]{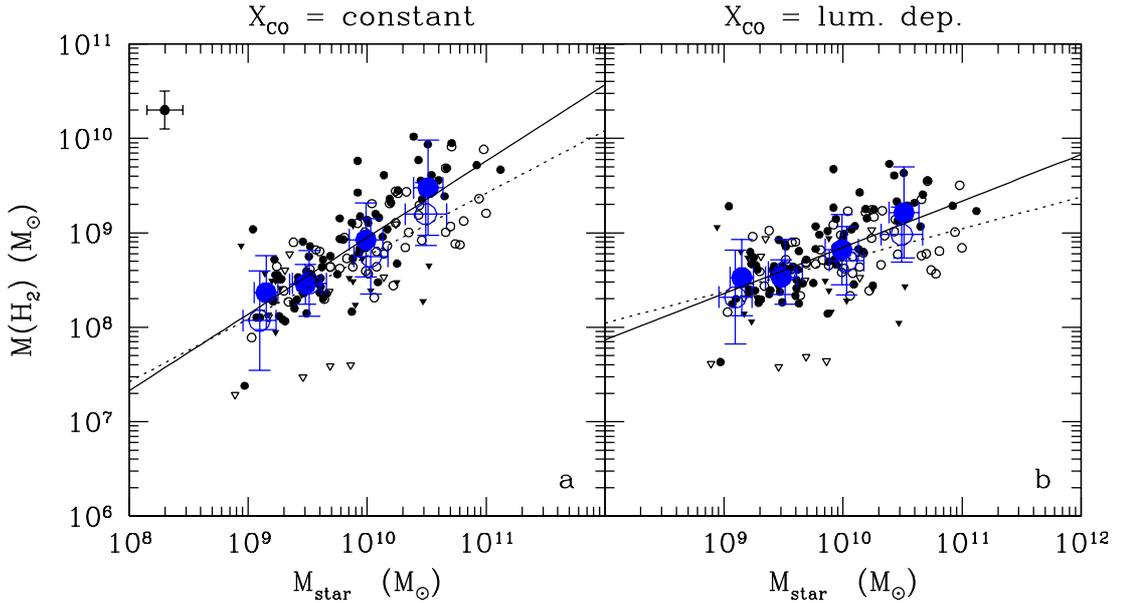}
   \caption{The relationship between molecular gas and stellar mass (in solar units)
   when the molecular gas mass is determined assuming a constant $X_{CO}$ factor (a, left) or the luminosity-dependent $X_{CO}$ factor 
   given in Boselli et al. (2002) (b, right). Symbols are as in Fig. \ref{calraggi}, with in addition big blue empty circles indicating mean values and standard deviations
   for H{\sc i}-deficient objects.
   The black solid line indicates the linear fit determined using all gas-rich galaxies, while the black dotted line that 
   determined for H{\sc i}-deficient objects (black empty symbols).  }
   \label{calmass}%
   \end{figure*}

To identify the scaling variable that minimises the scatter in the molecular gas mass vs. size/luminosity/mass relations, 
we plot in Fig. \ref{calraggi} and Fig. \ref{calmass} the relationship between the molecular gas mass and the $g$-band optical disc surface (defined as $\pi$$[r_{24.5}(g)]^2$)
and the stellar mass, respectively. Both figures have two panels, the left one where $M(H_2)$ is determined using a constant conversion factor, the right one
with a variable $X_{CO}$. 
As expected, the two variables are tightly correlated just because bigger galaxies have larger radii and more
gas content than smaller systems. The linear best fit to the data for H{\sc i}-rich 
spiral galaxies ($\mathrm{\hi}-def$ $\leq$ 0.4; direct fit)

\begin{equation}
{\log M(H_2) = c \times \log (Variable) + d}
\end{equation}

\noindent
gives the coefficients $c$ and $d$ listed in Table \ref{Tabcaldeffit}. The mean values in different bins of optical disc surface and stellar mass, indicated by blue large filled dots in both figures,
are listed in Table \ref{Tabcaldefdata}. Both fits and mean values are determined including CO non-detections at their upper limits. 
We choose to consider CO non-detections as detections because we want to avoid introducing any systematic bias in the determination of the best fit. The inclusion of upper limits 
as detections is justified by the fact that, because of the sample definition, all galaxies are $\sim$ at the same distance (15-25 Mpc). Since the CO surveys from which 
most of the data have been collected, and in particular our own which is the deepest among all, are limited to a given sensitivity (see paper I), the upper limits naturally 
follow in the low stellar and molecular gas 
mass range (lower left corner in Figs \ref{calraggi} and \ref{calmass}), and are mainly located below the fitted relations. 
The exclusion of these data from the fit might thus artificially flatten the fitted scaling relation, thereby introducing a systematic
bias in the results. Furthermore, if the cluster environment is responsible for gas removal, we would expect that, for a given sensitivity, the number of undetected objects increases in high-density
regions. A fair comparison of the statistical properties of cluster and field objects must thus consider non-detections.

   \begin{figure*}
   \centering
   \includegraphics[width=14cm]{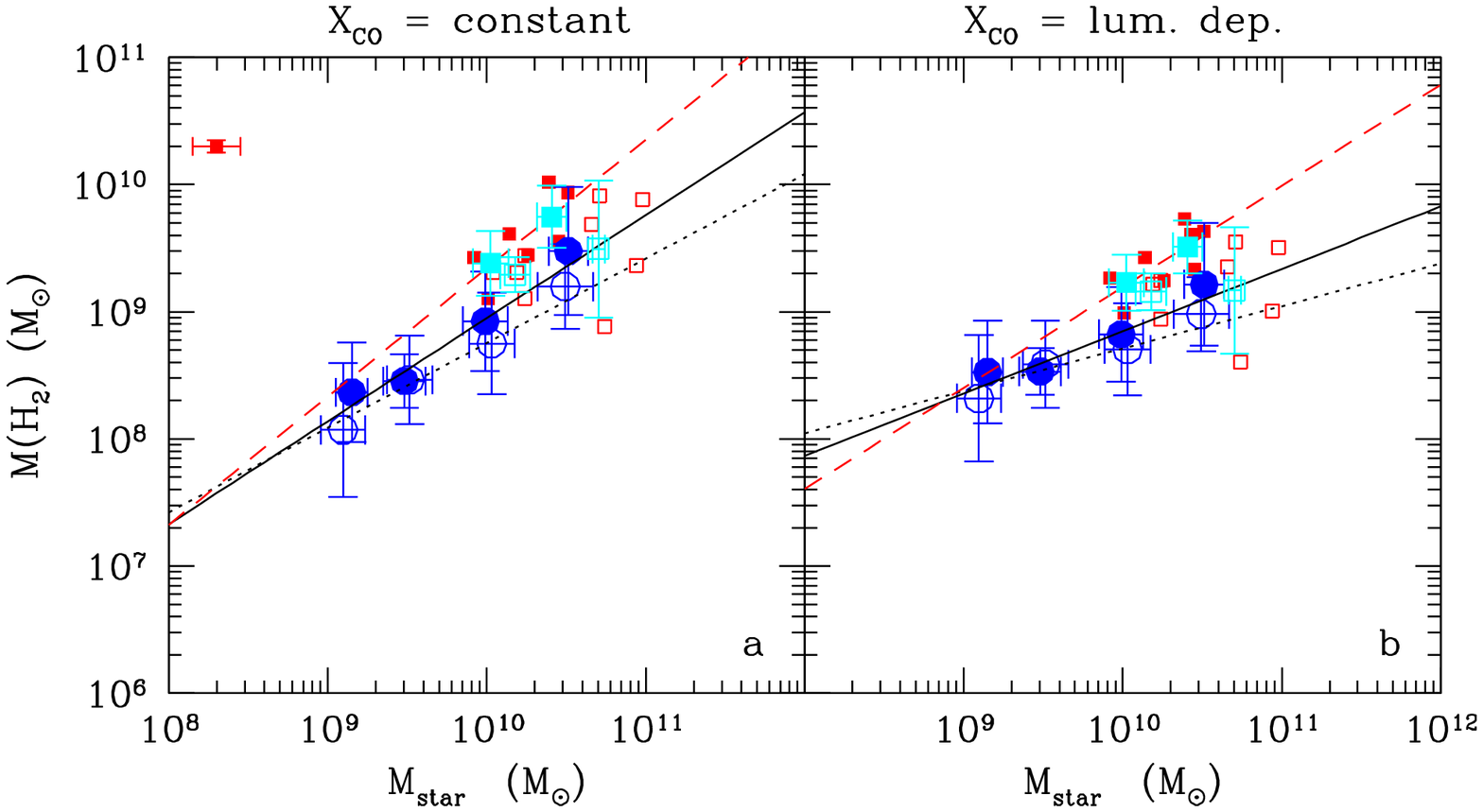}
   \caption{The relationship between molecular gas and stellar mass (in solar units)
   when the molecular gas mass is determined assuming a constant $X_{CO}$ factor (a, left) or the luminosity-dependent $X_{CO}$ factor 
   given in Boselli et al. (2002) (b, right) for the subsample of galaxies observed by Kuno et al. (2007). 
   Filled symbols indicate 
   objects with a normal H{\sc i} gas content ($\mathrm{\hi}-def$ $\leq$ 0.4), empty symbols H{\sc i}-deficient objects ($\mathrm{\hi}-def$ $>$ 0.4). 
   The typical error bars on the data is given in panel a. The big cyan filled squares 
   indicate the mean values for unperturbed systems ($\mathrm{\hi}-def$ $\leq$ 0.4), while the cyan empty squares 
   the mean values for H{\sc i}-deficient objects. The big blue filled and empty circles show the mean values determined 
   for the whole HRS sample for unperturbed and H{\sc i}-deficient objects, respectively
   The red dashed line indicates the linear fit determined using all gas-rich galaxies ($\mathrm{\hi}-def$ $\leq$ 0.4; red filled squares), 
   the black solid and dotted lines the best fit determined for the whole HRS sample for H{\sc i}-normal and H{\sc i}-deficient galaxies, respectively.}
   \label{calkuno}%
   \end{figure*}


Fig. \ref{calraggi}, \ref{calmass}, and Table \ref{Tabcaldeffit} all indicate that the physical parameter 
which minimises the dispersion in these two scaling relations is the total stellar mass. The dispersion in the various relations with stellar mass
are indeed a factor of $\sim$ 1.3 smaller than in those done using the optical area. 
This result only is in apparent contradiction with the work of Haynes \& Giovanelli (1984), who identified
the optical diameter as the best tracer for the atomic gas. Indeed the atomic gas has a much more extended distribution than the molecular phase, which has a steep radial gradient. It is thus not
surprising that the H{\sc i} content is more correlated to the outer extension of the stellar disc traced by the optical radius, while the molecular gas located in the inner regions
by the potential well of the galaxy, traced by the total stellar mass.
There is also another possible reason explaining the smaller dispersion in the $M(H_2)$ vs. $M_{star}$ with respect to the $M(H_2)$ vs. $\pi$$[r_{24.5}(g)]^2$ relation:
the stellar mass is determined from integrated magnitudes. Total magnitudes are less uncertain than optical isophotal diameters, which require the determination of the light radial profile
and are thus more sensitive to the sky subtraction and the determination of the fitting parameters (inclination, position angle, etc.). 

The reference scaling relation for the calibration of the gas deficiency parameter must use a scaling variable that is not affected by the interaction of
galaxies with the perturbing environment. Multizone chemo-spectrophotometric models of galaxy evolution especially tailored to 
mimic the perturbations induced by the environment on spiral discs 
have shown that, on relatively short timescales, ram pressure stripping events do not perturb the total stellar mass of galaxies (Boselli et al. 2006)\footnote{
They can, however, reduce the extension of the stellar disc when measured in photometric bands sensitive to the presence of young stellar populations. 
Even though the $g$-band is sensitive to the presence of relatively young stars, we did not observe any strong systematic difference in the isophotal radii
of isolated and cluster galaxies (Cortese et al. 2012a).}. There is a further reason to prefer the stellar mass to the optical area as scaling variable
to calibrate the molecular gas deficiency parameter. Molecular gas masses of most of the galaxies of our sample (78 \%) have been determined 
by correcting central single beam observations using an aperture correction depending on the optical diameter of the galaxies. Although this is a second order effect, it
might artificially reduce the scatter in the $M(H_2)$ vs. $\pi$$[r_{24.5}(g)]^2$ relation. 

Part of the scatter in the molecular gas mass vs. stellar mass relation might result from the uncertainty in the determination of the stellar mass through the use
of simple recipes as the one adopted in this work. To quantify this possible effect, we compare the stellar masses estimated following the prescription of Zibetti et al. (2009) used in this work
to those determined using two different methods. We first compare our stellar masses
to those determined using the recipe given in Boselli et al. (2009) based on $H$-band luminosities and $B-H$ colour-dependent stellar mass-to-light ratios.
These two prescriptions are totally independent since based on different sets of photometric data (SDSS for the former, Johnson the latter), stellar population synthesis models (the 2007 version of the Bruzual \&
Charlot 2003 models for Zibetti et al. 2009, Boissier \& Prantzos 2000 for Boselli
et al. 2009), IMF (Chabrier 2003 vs. Kroupa 2001). The two sets of stellar masses determined using these methods are consistent, with a typical scatter of 0.14 dex. We then compare these two sets of stellar masses 
to those determined by Boquien et al. (2012) using a SED fitting technique for a subsample of 64 HRS galaxies in common. For these objects stellar masses have been estimated using the CIGALE SED fitting 
code (Noll et al. 2009) based on the Maraston (2005) population synthesis models, and assuming a Kroupa (2001) IMF. The code has been run for a solar metallicity, and assuming two exponentially declining star formation
histories, the first one 13 Gyr old to reproduce the old stellar population, the second one recent to mimic the present star formation activity. Once again the stellar masses determined using the
prescription of Zibetti et al. (2009) and Boselli et al. (2009) are very consistent with those determined using the SED fitting technique. The comparison with both stellar masses has a typical scatter of 0.13 dex.
This test indicates that the uncertainty on the stellar mass of our galaxies is $\simeq$ 0.10 dex, and is independent of the method adopted to estimate it. The advantage of using several photometric bands 
in a SED fitting technique with respect to colour-based relations is balanced by the large uncertainty in the star formation history of these perturbed, cluster galaxies and by the use of a fixed, solar
metallicity (the HRS sample spans a relatively large range in metallicity, 8.4 $\lesssim$ 12+log O/H $\lesssim$ 8.8, Hughes et al. 2013). On the other hand, the colour-based stellar mass recipes 
such as those of Boselli et al. (2009) have been calibrated using chemo-photometric models of galaxy evolution able to reproduce the radial gradients of nearby galaxies and are thus perfectly 
tuned to reproduce the objects analysed in this work. The uncertainty on the stellar mass (0.10 dex) must be compared to that on the molecular gas mass, which is hard to quantify given the large uncertainty on the
$X_{CO}$ conversion factor that might change from galaxy to galaxy. A direct comparison can be done with the uncertainty on the total CO flux determination, which is $\simeq$ 12 \% ~(0.05 dex) in the subsample of galaxies 
with CO data from Kuno et al. (2007), and $\gtrsim$ 44 \% ~(0.16 dex) in the remaining objects. The uncertainty on the stellar mass only
dominates the scatter in the $M_{star}$ vs. $M(H_2)$ relation 
in the subsample of galaxies with high-quality CO maps from Kuno et al. (2007), but not on the whole sample, where the dispersion in the relation ($\sigma$ $\simeq$ 0.30) is significantly higher.


Identified the stellar mass as the best variable tracing the molecular gas content of spiral galaxies, we can also plot and calibrate the relationship between molecular gas and stellar mass
for the subsample of galaxies mapped by Kuno et al. (2007; Fig. \ref{calkuno}). 
Despite its poor statistical robustness (it is based on only eight galaxies with a normal atomic gas content), this calibration is well tuned to quantify the molecular gas
stripping of the subsample of galaxies homogeneously mapped in CO by Kuno et al. (2007).
This calibration should not be extended to low-mass systems given that the dynamic range covered by the subsample of galaxies with data from Kuno 
is limited to $M_{star}$ $\gtrsim$ 10$^{10}$ M$_{\odot}$. Figure \ref{calkuno}, however, shows that the extrapolation of the Kuno et al. calibration 
well matches the molecular gas vs. stellar mass relation observed in low-mass systems, where aperture corrections are expected to be more accurate.
The molecular gas mass of the mapped galaxies is systematically higher than that of similar 
galaxies with extrapolated data.
The systematic difference in the high stellar mass range might be partly due
to a larger uncertainty in the aperture correction in extended, massive systems than in small objects, where the beam of the telescope approximately matches the optical extension of the galaxy. 

The $c$ and $d$ coefficients given in Table \ref{Tabcaldeffit}, combined with eq. 1, can be used to estimate the expected molecular gas content $M(H_2)_{exp}$ of a 
galaxy of given stellar mass $M_{star}$. 
As in Boselli et al. (2002), we define a molecular hydrogen gas deficiency:

\begin{equation}
{H_2-def = \log M(H_2)_{exp} - \log M(H_2)_{obs} }
\end{equation}

\noindent
where the expected molecular gas mass of a target galaxy of stellar mass $M_{star}$, deduced from eq. 1 by substituting $M(H_2)$ with $M(H_2)_{exp}$, 
is compared to that determined from the CO observations
($M(H_2)_{obs}$). As defined, the molecular gas deficiency parameter is consistent with the atomic hydrogen deficiency parameter of Haynes \& Giovanelli
(1984) to quantify any possible effect on the atomic gas properties of galaxies in dense environments. Part of the dispersion in Fig. \ref{calmass}
can be correlated to other physical parameters. In paper II we have shown that the $M(H_2)/M_{star}$ ratio is indeed correlated to the specific
star formation rate, a parameter tightly connected to the cluster environment (Gavazzi et al. 2002).
Given the poor statistics in our subsample of H{\sc i}-normal objects, and in particular in those with high-quality integrated CO data (8 objects),
we cannot test whether the use of different coefficients for subsamples of galaxies selected according to different criteria (morphological type, effective 
surface brightness...) can reduce the scatter in the relation.
Equation 2 is in logarithmic scale, and thus indicates that a galaxy with an $H_2-def$ = 1 has 10 times less molecular hydrogen than a typical
unperturbed object of similar stellar mass.
The mean dispersion in the $M(H_2)$ vs. $M_{star}$ for galaxies with $\mathrm{\hi}-def$ $\leq$ 0.4 used to calibrate the $H_2$-deficiency parameter
is $\sigma$ = 0.27 and $\sigma$ = 0.31 for the whole sample (the first value is relative to a constant $X_{CO}$, the second one to a luminosity-dependent $X_{CO}$), and 
drops to $\sigma$ = 0.14-0.13 for the subsample of galaxies with integrated data. We can thus consider as molecular gas deficient objects those galaxies with $H_2-def$ $\gtrsim$ 0.3.
We recall, however, that both the H{\sc i}- and $H_2$-deficiency parameters should be considered only statistically. This is even more evident for $H_2-def$ where the uncertainties on the 
determination of the total CO luminosity due to aperture corrections are significantly more important than those on the estimate of the total H{\sc i} mass.

\subsection{Molecular gas stripping}

   \begin{figure*}
   \centering
   \includegraphics[width=14cm]{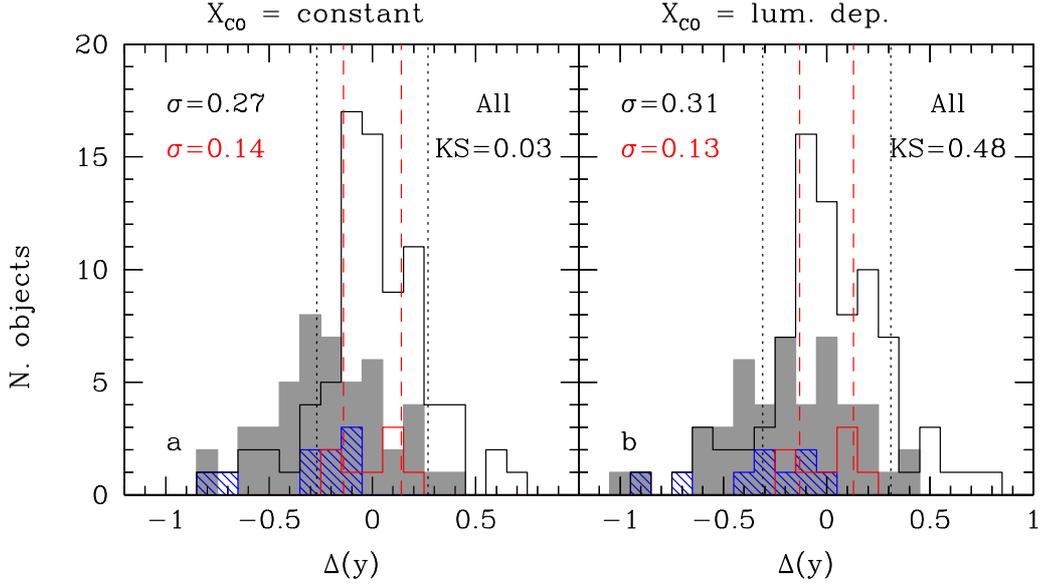}
   \caption{The distribution of the distance from the $M(H_2)$ vs. $M_{star}$ relations $\Delta(y)$ plotted in Figs. \ref{calmass} and \ref{calkuno} determined 
   assuming a constant $X_{CO}$ factor (panel a) or the luminosity-dependent $X_{CO}$ factor 
   given in Boselli et al. (2002) (panels). Negative values are for galaxies below the fitted relation.
   The black empty histogram gives the distribution of galaxies with $\mathrm{\hi}-def$ $\leq$ 0.3,
   the filled grey histogram that of the most H{\sc i}-deficient objects of the sample ($\mathrm{\hi}-def$ $>$ 0.6). Non-detected galaxies are included at their upper limits. 
   The red empty histogram shows
   the distribution of the spiral galaxies mapped in CO by Kuno et al. (2007) with $\mathrm{\hi}-def$ $\leq$ 0.4, the 
   blue dashed histogram that of the H{\sc i}-deficient galaxies ($\mathrm{\hi}-def$ $>$ 0.4). The black dotted and red dashed vertical lines give the one-sigma dispersion in the distribution
   determined for the whole sample of H{\sc i}-normal ($\mathrm{\hi}-def$ $\leq$ 0.4) galaxies (black) and for the subsample of objects mapped by Kuno et al. (2007; red). 
   $\sigma$ indicates the mean dispersion of the different distributions of unperturbed galaxies. KS indicates the probability that the two
   galaxy populations are driven by the same parent population (Kolmogorov-Smirnov test). }
   \label{dispersione}%
   \end{figure*}

   \begin{figure*}
   \centering
   \includegraphics[width=14cm]{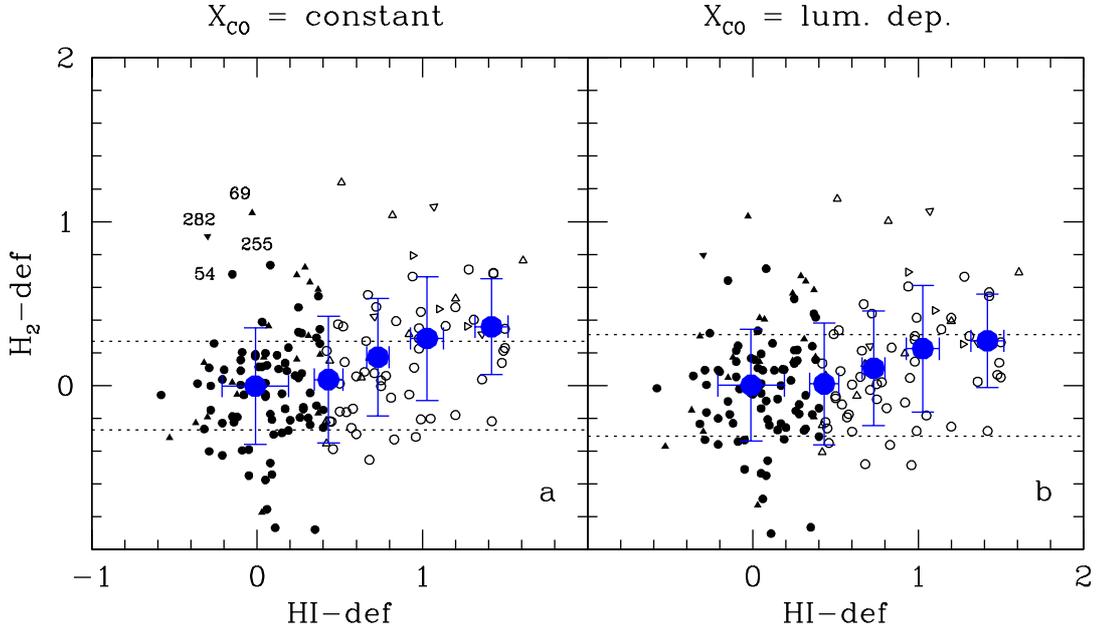}
   \caption{The relationship between molecular gas and atomic gas deficiency for spiral galaxies
   when the molecular gas mass is determined assuming a constant $X_{CO}$ factor (panel a) or the luminosity-dependent $X_{CO}$ factor 
   given in Boselli et al. (2002) (panel b). Symbols are as in Fig. \ref{calraggi}.
   The horizontal dotted lines show the one-sigma dispersion of $H_2$-deficiency parameter as determined from the calibrating relations given in Fig. \ref{calmass}.}
   \label{defHIdefH2}%
   \end{figure*}


   \begin{figure*}
   \centering
   \includegraphics[width=14cm]{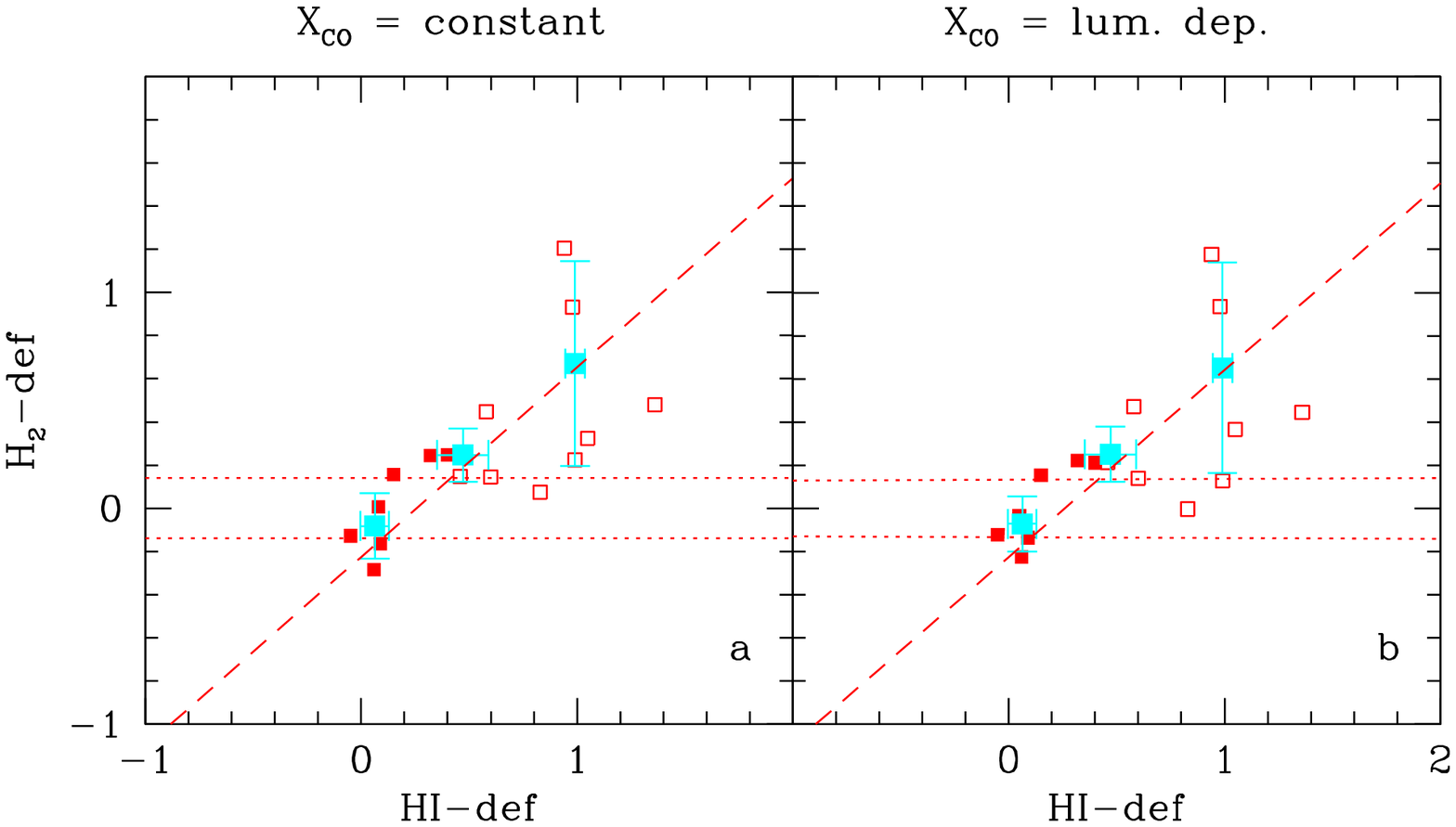}
   \caption{The relationship between molecular gas and atomic gas deficiency for spiral galaxies mapped in CO by Kuno et al. (2007) (red symbols)
   when the molecular gas mass is determined assuming a constant $X_{CO}$ factor (panel a) or the luminosity-dependent $X_{CO}$ factor 
   given in Boselli et al. (2002) (panel b). Symbols are as in Fig. \ref{calkuno}. The red dashed line shows the bisector fit (the parameters 
   of the fit are given in Table \ref{Tabdefdeffit}).
   The horizontal dotted lines show the one-sigma dispersion of $H_2$-deficiency parameter as determined from the calibrating relations given in Fig. \ref{calkuno}.}
   \label{defHIdefH2kuno}%
   \end{figure*}

The effects of the cluster environment on the molecular gas content of all the HRS galaxies can already be noticed from the analysis of Fig. \ref{calmass} where,
for a given stellar mass, H{\sc i}-deficient cluster galaxies ($\mathrm{\hi}-def$ $>$ 0.4; empty symbols) are located systematically below unperturbed objects ($\mathrm{\hi}-def$ $\leq$ 0.4; filled symbols).
The same effect, if present, is very marginal in low-mass systems. The best fit of the $M(H_2)$ vs. $M_{star}$ relation determined using all unperturbed galaxies (black solid line)
is indeed steeper and higher than that measured using all the H{\sc i}-deficient galaxies (black dotted line). The two best fits match at $M_{star}$ $\simeq$ 10$^9$ M$_{\odot}$, 
the lower limit in stellar mass sampled by our data. The different behaviour of massive and dwarf systems might be related to a bias in the CO data. At low stellar masses, where the 
sample is not complete, there is a systematic lack of CO data in the most HI-deficient galaxies of the sample, as discussed in sect. 4.3. The mean value of $M(H_2)$ determined 
for the low-mass systems should thus be considered as an upper limit. 

To quantify statistically this effect we show in Fig. \ref{dispersione} the distribution of the distance $\Delta(y)$ from the 
fitted $M(H_2)$ vs. $M_{star}$ relations shown in Figs. \ref{calmass} and \ref{calkuno}
for different subsamples of galaxies. Since we expect that the molecular gas content of galaxies decreases gradually with their degree of perturbation,
thus with their H{\sc i}-deficiency parameter, we compare the distribution of the distance from the fitted relations for two well distinct populations.
We identify unperturbed objects those with $\mathrm{\hi}-def$ $\leq$ 0.3. This is a more stringent limit than the one generally used in this work. We recall that 
$\mathrm{\hi}-def$ = 0.3 corresponds to the dispersion in the H{\sc i}-deficiency distribution of isolated galaxies first determined by Haynes \& Giovanelli (1984). We compare their distribution
to that of the most H{\sc i}-deficient objects of the sample ($\mathrm{\hi}-def$ $>$ 0.6). 
Figure \ref{dispersione} shows that the most H{\sc i}-deficient galaxies of the sample have, on average, a lower molecular gas content per unit stellar mass than unperturbed objects.
The mean value for the 80 galaxies with $\mathrm{\hi}-def$ $\leq$ 0.3 is $\Delta(y)$ = 0.00 $\pm$ 0.03 (error on the mean) for both a constant or a luminosity-dependent $X_{CO}$, 
while that of the 47 most H{\sc i}-deficient objects of the sample ($\mathrm{\hi}-def$ $>$ 0.6) is $\Delta(y)$ = -0.21 $\pm$ 0.04 ($X_{CO}$ constant) and $\Delta(y)$ = -0.18 $\pm$ 0.04 ($X_{CO}$
luminosity-dependent).
A Kolmogorov-Smirnov test indicates that the probability that the two distributions are drawn by the same underlying population is less than 0.5\%, and this regardless the use of a constant 
or a luminosity-dependent $X_{CO}$ conversion factor\footnote{The mean values of H{\sc i}-normal and H{\sc i}-deficient galaxies when a unique threshold of $\mathrm{\hi}-def$ $=$ 0.4 is
used to separate the two subsamples are $\Delta(y)$ = -0.01 $\pm$ 0.03 for both a constant or a luminosity-dependent $X_{CO}$, 
and $\Delta(y)$ = -0.15 $\pm$ 0.04 ($X_{CO}$ constant) and $\Delta(y)$ = -0.12 $\pm$ 0.04 ($X_{CO}$ luminosity-dependent). Using this definition to identify perturbed and unperturbed
objects, the probability that the two distributions are drawn by the same underlying population is still less than 0.5\% when a constant $X_{CO}$ is assumed, and increases to 9.1\% 
for $X_{CO}$ luminosity-dependent.}. 
The same systematic difference is also observed in the distribution of $\Delta(y)$ of H{\sc i}-normal ($\mathrm{\hi}-def$ $\leq$ 0.4)
and H{\sc i}-deficient ($\mathrm{\hi}-def$ $>$ 0.4) galaxies with high-quality data from Kuno et al. (2007), or for the subsample of massive ($M_{star}$ $>$ 10$^{10}$ M$_{\odot}$) objects  
(the number of objects belonging to these subsamples is unfortunately too small to make reliable statistical tests such as the Kolmogorov-Smirnov test).
\\

To quantify the effect of the Virgo cluster environment on the molecular gas content of the HRS galaxies, we plot in Figs. \ref{defHIdefH2} 
the relationship between the H$_2$ ($H_2-def$) and the H{\sc i} ($\mathrm{\hi}-def$) deficiency parameters for the whole sample of galaxies. 
Given that the H{\sc i}
gas is easily removed in any kind of gravitational or dynamical interaction with the cluster environment (Boselli \& Gavazzi 2006), the $\mathrm{\hi}-def$ parameter 
is here used as a proxy for the induced perturbation. Figure \ref{defHIdefH2} shows a dispersed but statistically significant relation between the two 
variables. This relation seems robust vs. the adoption of a constant or a luminosity-dependent $X_{CO}$ conversion factor (left vs. right panels), 
with a probability $P$ that the two variables are correlated always larger than 99 \%. 
Figure \ref{defHIdefH2} shows that, on average, H{\sc i}-poor galaxies have a lower molecular gas content than unperturbed objects. The observed relation is steeper 
for the unbiased subsample of massive galaxies than for the whole HRS sample.
The lack of moleculr gas, however, seems
significantly less important than that of atomic gas: the mean value of $H_2-def$ for all the H{\sc i}-normal ($\mathrm{\hi}-def$ $<$ 0.4) 
HRS spiral galaxies of the sample is  $H_2-def$ $\simeq$ 0.00 $\pm$ 0.04 (error on the mean) and increases up to $H_2-def$ $\simeq$ 0.3$\pm$ 0.08  for the most H{\sc i}-poor 
objects ($\mathrm{\hi}-def$ $\geq$ 1.2) (see Table \ref{Tabdefdefdata}). Limited to massive galaxies, the $H_2$-deficiency parameter increases by $\simeq$ 0.4
from the unperturbed to the most H{\sc i}-deficient objects of the sample ($\sim$ 4 $\sigma$ effect).

As discussed in Fumagalli et al. (2009), 
there are several H{\sc i}-deficient galaxies with a normal molecular gas content. 
At the same time, Fig. \ref{defHIdefH2} 
shows also the presence of a few spiral galaxies with a normal atomic gas content but deficient in molecular hydrogen. These objects are identified in panel a of Fig. \ref{defHIdefH2}, and are the galaxies 
HRS 54 (NGC 3681), HRS 69 (NGC 3898), HRS 255 (NGC 4688) and HRS 282 (NGC 4803). Clearly the uncertainty on the estimate of the total CO emission of HRS 69 and HRS 255 is very high given the filling factor of the telescope 
beam during the observations ($ff$ = 0.107 in HRS 69, $ff$=0.036 in HRS 255, see paper I). We also notice that the UV (Cortese et al. 2012a), H$\alpha$ (Boselli et al., in prep.)
or mid-infrared (Bendo et al. 2012) images of HRS 54 and 69 are characterised by an external ring of HII regions, thereby suggesting that the molecular gas associated to these star forming regions 
might not be detected by the central beam observation. We recall that HRS 282 is morphologically identified as "compact", and is much more similar to a quiescent, early-type galaxy than 
to a star forming spiral. It is indeed one of the few bright spirals non-detected by SPIRE (Ciesla et al. 2012). 

The trend between the H{\sc i}- and H$_2$-deficiency parameters is very weak, with the most H{\sc i}-poor galaxies only having, on average, a factor of $\simeq$ 2 less molecular gas (at 4 $\sigma$ level) 
than unperturbed objects. This small variation is hardly detectable with the whole sample of galaxies given that it is just a factor of $\simeq$ 2 the typical uncertainty on the 
molecular gas mass estimate, and can thus be taken only with a statistical sense. This weak trend was not detectable with the data set of Boselli et al. (2002), where the typical dispersion 
in the H$_2$-deficiency parameter of unperturbed objects was as large as $\simeq$ 0.5, nor with the very small sample of galaxies in A1367 by Scott et al. (2013).   

We can check whether the trends observed using the whole HRS sample are also shared by the subsample of galaxies with high-quality CO data from Kuno et al. (2007). As for Fig. \ref{calmass}, Fig.
\ref{calkuno} shows a systematic difference in the molecular gas mass vs. stellar mass relation in the CO mapped galaxies. Indeed those objects with a normal atomic gas content have systematically higher 
molecular gas masses than the perturbed objects ($\mathrm{\hi}-def$ $\ge$ 0.4). The small dispersion in the $M(H_2)$ vs. $M_{star}$ relation, probably due to the small uncertainty in the 
molecular gas determination, indicates that this difference is significant. We can also see the relationship between the H{\sc i}- and H$_2$-deficiency parameters in Fig. \ref{defHIdefH2kuno}
when the H$_2$-deficiency parameter of the Kuno et al. galaxies is calculated using the calibration given in Fig. \ref{calkuno}. Again, Fig. \ref{defHIdefH2kuno} shows a well defined trend between the 
two variables ($P$ $>$ 99.9 \%), thereby confirming the results obtained for the whole sample (the results of the bisector fit
are given in Table \ref{Tabdefdeffit}). Here the dispersion in the H$_2$-deficiency parameter of unperturbed galaxies is only $\simeq$ 0.14, thus a factor of $\sim$
4 less than the mean H$_2$-deficiency of the most H{\sc i}-perturbed galaxies of the Kuno et al. sample.
The same H$_2$ vs. H{\sc i}-deficiency relations for the whole sample
of HRS galaxies but determined using the H$_2$-deficiency calibration of Kuno et al. give steeper  
and stronger ($P$ $>$ 99.9\%) trends between the two variables than those shown in Fig. \ref{defHIdefH2}.
The slope in the $H_2-def$ vs. $\mathrm{\hi}-def$ relation is $<$ 1 for every H$_2$-deficiency calibration and analysed subsample, 
whcih suggests that the molecular gas depletion is less efficient than the atomic one.\\

The interpretation of this new, observational evidence, i.e. the molecular gas deficiency in cluster galaxies, can follow two different arguments. The first one is related to the transformation of H{\sc i} to H$_2$,
the second one ram pressure stripping of the molecular gas (Fumagalli et al. 2009).

\subsubsection{Molecular gas formation}

Following the theoretical
motivations of Krumholz et al. (2008, 2009), the molecular gas phase is formed from the condensation of the atomic hydrogen, which only happens when the H{\sc i} column density exceeds a
critical value of $\Sigma(gas)$ $\sim$ 10 M$_{\odot}$ pc$^{-2}$. The decrease of the atomic gas column density resulting from 
the gas stripping in cluster galaxies, probably related to a ram pressure phenomenon (Boselli \& Gavazzi 2006), drops the H{\sc i} column density to 
values of $\Sigma(gas)$ $\lesssim$ 2 M$\odot$ pc$^{-2}$ (Cayatte et al. 1994, Chung et al. 2009a, Fumagalli et al. 2009), thus below this
critical density. This effect might induce a lower formation rate of the molecular gas phase, producing thus molecular gas poor cluster galaxies.
The timescale for molecular gas formation on dust grains is indeed quite short, of the order of 10-30 Myr (Liszt 2007; Goldsmith et al. 2007; Lee et al. 2012).
This phenomenon, however, would not affect the dust component associated to the molecular gas and thus does not explain the observed truncation of the far-infrared radial profiles
of H{\sc i}-deficient late-type galaxies in the Virgo cluster (Cortese et al. 2010; 2012b).\\

\subsubsection{Ram pressure stripping}

In the assumption that the radial molecular hydrogen distribution 
linearly follows the CO emission, i.e. that the radial variations of the $X_{CO}$ conversion factor due to metallicity gradients are minor 
compared to the variations of the CO emission (Sandstrom et al. 2013), the analysis presented in this work just indicates that the molecular gas phase is removed  by a ram pressure stripping event less efficiently than the
atomic one. This can be understood considering that the CO emission over the disc of galaxies is exponentially declining with a characteristic disc scalelength
$r_{CO}$ $\sim$ 0.2 the optical isophotal radius (Lisenfeld et al. 2011 and references therein), while the atomic hydrogen has a much
flatter and extended distribution (Cayatte et al. 1994, Bigiel \& Blitz 2012). The atomic gas phase,
which is dominating the gas phase in the outer disc, is thus the one less anchored
to the gravitational potential well of the galaxy. It can be easily removed during any kind of gravitational interaction with nearby companions or with the cluster gravitational potential well
(galaxy harassment), or during the dynamical interaction of the galaxy with the hot and dense intracluster medium (Boselli \& Gavazzi 2006). The molecular phase,
more centrally peaked, thus trapped within the gravitational potential well of the galaxy, is less efficiently removed during this kind of interaction. \\

The ram pressure stripping scenario is supported by
recent \textit{Herschel} observations of the HRS. Cortese et al. (2010, 2012b)
have shown that the dust component, tightly associated to the molecular gas phase of the ISM, is stripped in the most H{\sc i}-deficient Virgo cluster galaxies, which are indeed the same
galaxies showing molecular gas deficiency. This result is also in agreement with the observation of targeted objects with evidence of ongoing molecular gas stripping, such as
NGC 4522, NGC 4438, and NGC 4330 in Virgo (Vollmer et al. 2008a, 2009, 2012a) or ESO 137011 in Abell 3627 (Sivanandam et al. 2010).
We also notice that, among the brightest galaxies mapped by Kuno and collaborators, those with the highest H$_2$-deficiencies have all truncated 
(NGC 4419, 4501, 4569, 4579) or anemic (NGC 4548) H$\alpha$ (Koopmann \& Kenney 2004) and H{\sc i} (NGC 4548, 4569, 4579) profiles (Cayatte et al. 1994), 
typical signs of a recent interaction with the intracluster medium. The detailed analysis of some of these objects have indeed shown 
an ongoing (NGC 4501, Vollmer et al. 2008b) or recent (NGC 4548, Vollmer et al. 1999; NGC 4569, Vollmer et al. 2004, Boselli et al. 2006)
ram pressure stripping event.\\

\begin{figure}
   \centering
   \includegraphics[width=8cm]{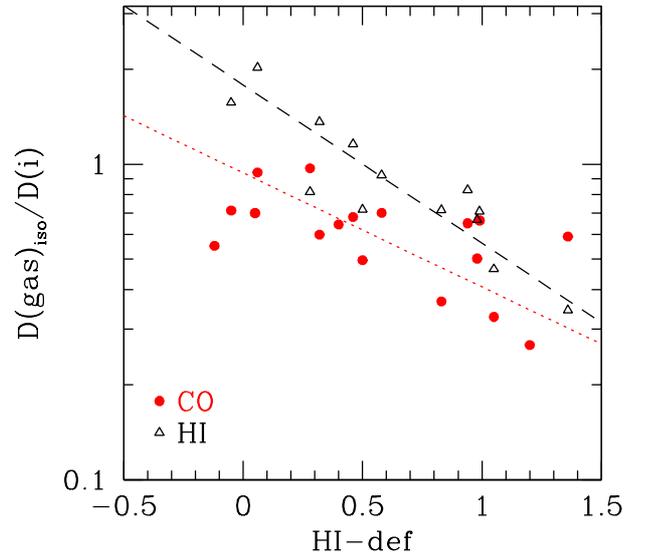}
   \caption{The relationship between the gas-to-stellar ($i$-band) isophotal diameter ratio and the H{\sc i}-deficiency parameter. Red filled dots are for 
   CO data, black open triangles for H{\sc i} data. The red dotted and black dashed lines indicate the best fit to the molecular 
   and atomic 
   gas data, respectively (bisector fit).}
   \label{risolte}%
   \end{figure}

To test whether this scenario is consistent with the data, we plot in Fig. \ref{risolte} the relationship between the ratio of the atomic and molecular gas-to-stellar ($i$-band) 
isophotal diameter ratio and the H{\sc i}-deficiency parameter. CO and H{\sc i} isophotal diameters are taken from the subsample of HRS galaxies mapped by Chung et al. (2009b)\footnote{The galaxy NGC 4536 (HRS 205) is not included 
in the plot because not fully mapped in CO by Chung et al. (2009b)} in CO and by Chung et al.
(2009a) in H{\sc i}. Both molecular and atomic gas isophotal diameters are determined where the gas column density drops to 1 M$_{\odot}$ pc$^{-2}$. 
The figure shows two statistically significant anticorrelations between the gas (atomic and molecular)-to-stellar isophotal 
diameter ratios and the H{\sc i}-deficiency parameter. A bisector fit to the data gives
log $\frac{D(\mathrm{\hi})}{D(i)}$ = -0.50 $\times$ $\mathrm{\hi}-def$ +0.25, $\rho$ -0.80, $P>$ 99.5\% and
log $\frac{D(CO)}{D(i)}$ = -0.36 $\times$ $\mathrm{\hi}-def$ -0.03, $\rho$ -0.57, $P>$ 99\%, thereby suggesting that gas erosion in cluster galaxies is an outside-in process.
The atomic gas and, to a lower extent, the molecular gas, are removed in the outer parts of galaxies belonging to high-density environments. 
A steeper relation between $D(\mathrm{\hi})/D(i)$ vs. $\mathrm{\hi}-def$ than between $D(CO)/D(i)$ vs. $\mathrm{\hi}-def$ is consistent with our previous 
result that the slope of the $H_2$- vs. H{\sc i}-deficiency relation is smaller than one (see Figs. \ref{defHIdefH2}-\ref{defHIdefH2kuno}). This result is also
perfectly consistent with a ram pressure stripping scenario, where only the gaseous component is perturbed during the interaction. Gravitational interactions (harassment) would indeed perturb also the
stellar component. We recall that the $D(CO)/D(i)$ vs. $\mathrm{\hi}-def$ relation shown in Fig. \ref{risolte} also perfectly matches the observed cold dust-to-optical 
isophotal diameter ratio vs. H{\sc i}-deficiency relation determined using \textit{SPIRE/Herschel} data for the HRS galaxies by Cortese et al. (2010).
\\

Recent high-resolution (40 pc) 3D hydrodynamical simulations of galaxies undergoing a ram pressure stripping event support this scenario. Tonnesen \& Bryan (2009) have included in their code radiative cooling on
a multiphase medium. This naturally produces a clumpy ISM with densities spanning six orders of magnitude, thus quite representative of the physical conditions encountered in normal, late-type galaxies. Their 
simulations show that under these conditions the gas is stripped more efficiently up to the inner regions with respect to an homogeneous gas. They also show that all the low density, diffuse gas is quickly stripped    
at all radii. When the ram pressure stripping is strong there is also less gas at high densities. The deficiency in high-density regions results from the lack of the diffuse component feeding giant molecular clouds.
This evidence suggests that the ram pressure stripping process, at the origin of the H{\sc i}-deficiency observed in cluster galaxy, is able to remove also the molecular gas. For this reason we interpret the trend between the H$_2$ and the
H{\sc i}-deficiency parameters shown in Fig. \ref{defHIdefH2} as an evidence of molecular gas ram pressure stripping in cluster galaxies.

\subsection{Implications on the star formation history of cluster galaxies}

   \begin{figure*}
   \centering
   \includegraphics[width=14cm]{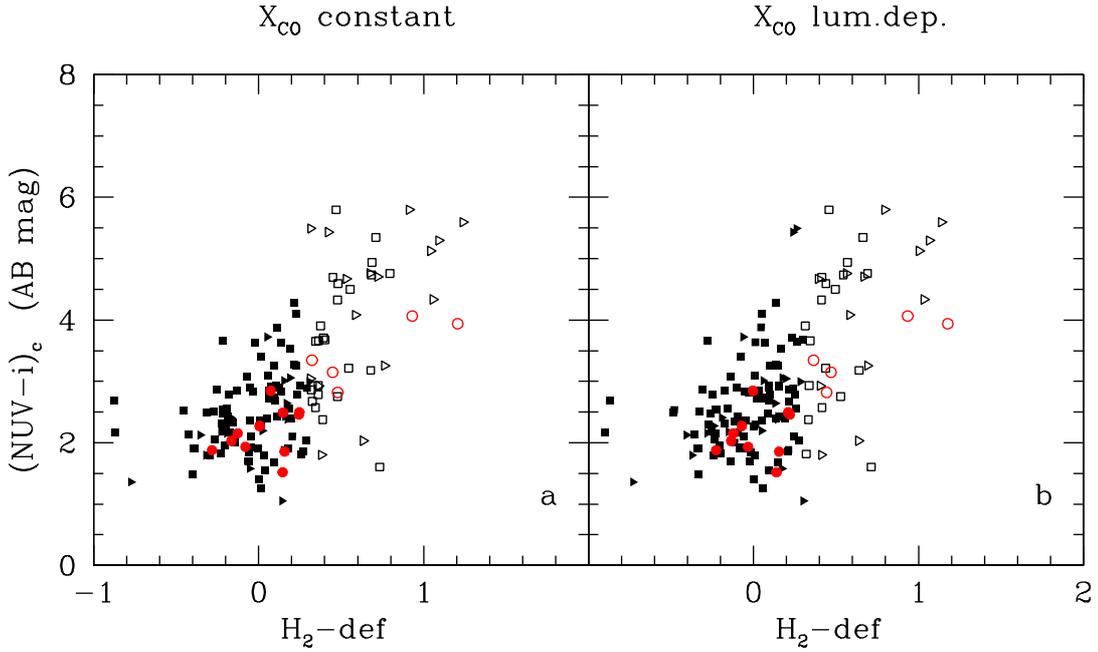}
   \caption{The NUV-$i$ vs. H$_2$-deficiency relation for the whole HRS galaxy sample. NUV-$i$ colours (in AB system) are corrected for Galactic and internal dust attenuation
   following Hao et al. (2011). Galaxies are coded according to their H$_2$-deficiency parameter,
   where the $H_2-def$ has been measured using a constant (panel a) or a luminosity-dependent (panel b) $X_{CO}$ conversion factor. Black squares indicate spiral galaxies
   whose molecular gas content has been determined using aperture corrections, red dots those objects with available CO integrated maps from Kuno et al. (2007).
   Filled symbols are for galaxies with $H_2-def$ $\leq$ 0.3, open symbols for objects with $H_2-def$ $>$ 0.3. For the black squares, the H$_2$-deficiency parameter has been 
   determined using the $M(H_2)$ vs. $M_{star}$ calibration based on all galaxies (lines 3 and 4 in Table \ref{Tabcaldeffit}), for the red symbols the calibration done on the subsample of
   galaxies with mapped observations from Kuno et al. (2007) (lines 5 and 6 in Table \ref{Tabcaldeffit}).
   }
   \label{NUVidef}%
   \end{figure*}

   \begin{figure*}
   \centering
   \includegraphics[width=18cm]{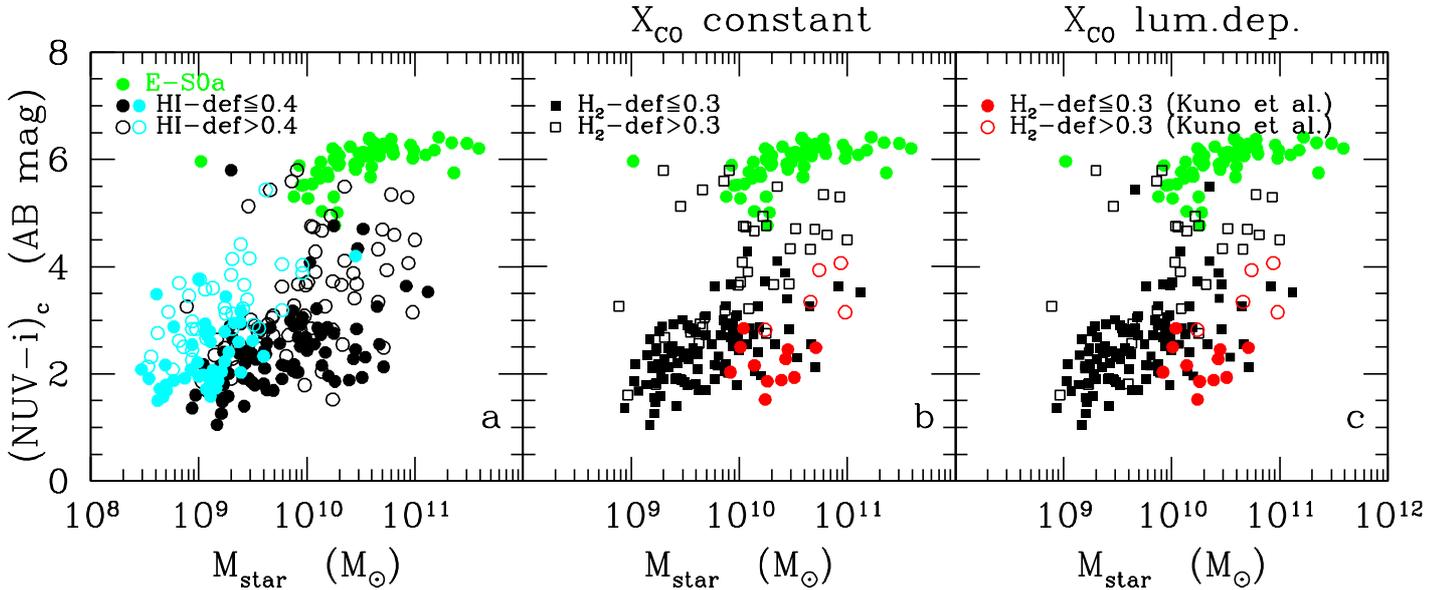}
   \caption{The NUV-$i$ vs. stellar mass relation for the whole HRS galaxy sample. NUV-$i$ colours (in AB system) are corrected for Galactic and internal dust attenuation
   following Hao et al. (2011). In all panels, green filled dots indicate early-type galaxies (E-S0a). In panel a spiral galaxies are coded according to their H{\sc i}-deficiency parameter
   (filled dots for $\mathrm{\hi}-def$ $\leq$ 0.4; open circles for $\mathrm{\hi}-def$ $>$ 0.4). Black symbols indicate galaxies with available CO data, cyan symbols galaxies 
   with H{\sc i} data but never observed in CO (thus lacking in panels b and c). In panels b and c spiral galaxies are coded according to their H$_2$-deficiency parameter,
   where the $H_2-def$ has been measured using a constant (panel b) or a luminosity-dependent (panel c) $X_{CO}$ conversion factor. Black squares indicate spiral galaxies
   whose molecular gas content has been determined using aperture corrections, red dots those objects with available CO integrated maps from Kuno et al. (2007).
   Filled symbols are for galaxies with $H_2-def$ $\leq$ 0.3, open symbols for objects with $H_2-def$ $>$ 0.3. For the black squares, the H$_2$-deficiency parameter has been 
   determined using the $M(H_2)$ vs. $M_{star}$ calibration based on all galaxies (lines 3 and 4 in Table \ref{Tabcaldeffit}), for the red symbols the calibration done on the subsample of
   galaxies with mapped observations from Kuno et al. (2007) (lines 5 and 6 in Table \ref{Tabcaldeffit}).
   }
   \label{CMR}%
   \end{figure*}

To study the physical consequences of the molecular gas stripping of cluster galaxies on the process of star formation, we plot in
Fig. \ref{NUVidef} the relationship between the $NUV-i$ colour index, here taken as a proxy for the specific star formation rate, and the H$_2$-deficiency parameter. 
\footnote{NUV magnitudes are corrected for internal dust attenuation according to Hao et al. (2011), while the attenuation in the $i$-band is simply determined by scaling those in the NUV band 
using the Galactic extinction law of Fitzpatrick \& Massa (2007). They are also corrected for Galactic extinction according to Schlegel et al. (1998)}. 
There is a clear trend between the two variables regardless of the adopted $H_2-def$  definition which indicates that $H_2$-deficient spiral galaxies have, on average, redder colours 
than objects with a normal molecular gas content. The lack of molecular gas quenches the activity of star formation making galaxies redder. This evolutionary picture can be seen
also in a $NUV-i$ vs. $M_{star}$ colour-magnitude relation (Fig. \ref{CMR}).
The atomic and molecular gas-deficient cluster spiral galaxies of the sample are located in the "green valley" consistently with the picture where they are
migrating from the blue cloud to 
the red sequence (Hughes \& Cortese 2009; Cortese \& Hughes 2009; Gavazzi et al. 2013a,b). Figure \ref{CMR} also shows that atomic and molecular gas
deficient objects in the green valley are mainly massive galaxies ($M_{star}$ $\ge$ 10$^{10}$ M$_{\odot}$). There is a lack of H$_2$-deficient objects in the low stellar mass range. 
This clearly results from a selection bias in the sample since at low stellar masses there
is a large number of H{\sc i}-deficient galaxies, mainly located in the "green valley", without CO data. This bias explains the 
trend observed in Fig. \ref{calmass}, where the few H{\sc i}-deficient low-mass galaxies have molecular gas contents comparable to that of unperturbed systems. \\
Our data show that the gas consumption timescales of cluster and isolated objects are comparable (see Table
\ref{Tabtau}).  
Because of the tight relation between gas column density and star
formation, the stripped gas does not further supply the star formation process reducing the global activity  
in the perturbed galaxies. The ratio of the gas mass to the star formation activity, or in other words the gas consumption timescale, thus remains fairly constant and
similar to that of unperturbed objects. This result is consistent with the pixel per pixel analysis of resolved, Virgo cluster galaxies done by Vollmer et al. (2012b).\\

What is the physical process governing the future evolution of these galaxies?
To answer this question we first compare the gas depletion timescales determined in this work to other representative timescales for the evolution of galaxies in clusters.
The typical crossing of a cluster such as Virgo is $\sim$ 1.7 Gyr (Boselli \& Gavazzi 2006), while
the timescale necessary to strip gas from a galaxy via ram pressure is of the order of $\sim$ 100-200 Myr 
(Vollmer et al. 2004, 2008a, 2009, 2012a; Boselli et al. 2006; Roediger \& Bruggen 2007; Crowl \& Kenney 2008). This stripping process only removes a 
fraction of the total gas from massive objects, while it has dramatic effects in dwarfs,
where the star formation process is completely stopped (Boselli et al. 2008a). 
The simulations of Tonnesen et al. (2007) done considering only the atomic, diffuse gas phase indicate that ram pressure is the dominant process responsible for gas sweeping up to the virial radius of the cluster,
which in Virgo is of 1.68 Mpc. Models and simulations indeed show that, although dominant in the core of the cluster
where the velocity of the galaxy and the density of the ICM are maximal, ram pressure stripping is an ongoing process eroding the gaseous component all over the orbit of the galaxy 
within the cluster (Roediger \& Hensler 2005; Roediger \& Bruggen 2006, 2007). 
If most of the gas is removed rapidly, the total gas sweeping requires longer timescales. The simulations of Tonnesen et al. (2007) done using a single phase component of the ISM
show that the gas is totally removed on timescales $\geq$ 1 Gyr. Timescales of the order of 1.5 Gyr are also obtained using 3D hydrodynamical simulations by Roediger \& Bruggen (2007) for a substantial reduction of the total gas content 
of the perturbed galaxy via ram pressure. 
These hydrodynamical simulations based on a single and homogeneous gas phase for the ISM, however, generally underestimate the efficiency of ram pressure stripping (Tonnesen \& Bryan 2009).
Multiphase hydrodynamical simulations indicate that gas ablation in all its phases (from diffuse atomic to dense molecular gas) can take place at all galactic radii if ram pressure stripping
is sufficiently strong, as indeed is the case in rich clusters of galaxies such as Virgo (Tonnesen \& Bryan 2009). There is also observational evidence indicating that ram pressure stripping 
is efficient up to the cluster virial radius (Scott et al. 2012; Gavazzi et al. 2001;
Yagi et al. 2010; Fossati et al. 2012).
It is thus reasonable to assume that ram pressure stripping can completely remove the gas of the ISM on
timescales of $\lesssim$ 1.5 Gyr in galaxies within the virial radius\footnote{This assumption is different from the one taken by Cortese \& Hughes (2009) based on earlier
simulations, where ram pressure stripping is not able to remove the whole gas content of massive spiral galaxies in rich clusters.}. 

How does this timescale compare with the timescale for gas consumption via star formation (starvation)? We recall that the definition of starvation that we adopt here 
is the consumption of the gas via star formation when the cosmological infall of pristine gas is stopped (Boselli et al. 2006). This definition is quite different from the original one given in Larson et al. (1980), 
who considered in the starvation process also the removal of the gas rich envelopes of galaxies. Here we consider any process able to remove any gaseous component associated to the galaxy, including the H{\sc i}
gas located well outside the optical disc, as ram pressure stripping. 
To estimate the gas consumption timescales we should consider the contribution of the recycled gas.
The amount of recycled gas during the whole life of galaxies is generally assumed of the order of $\sim$ 30\% of their stellar mass (Kennicutt 1994). Since a large fraction of this gas, the one produced by the oldest stars, 
has been already injected into the ISM before the first interaction with the ICM and partly removed by the ram pressure stripping event, we have to consider here only the fraction of recycled gas
produced after the beginning of the interaction. The chemo-spectrophotometric models expressly tailored to take the perturbation induced by the cluster environment on galaxy evolution
that we have developed into account
can be used for this purpose (Boselli et al. 2008a). They indicate that the amount of recycled gas after a ram pressure stripping event is of the order of 9.0 10$^8$ M$_\odot$ for a 
massive galaxy of $M_{star}$ = 4.6 10$^{10}$ M$_\odot$, and of 4.2 10$^7$ M$_\odot$ for an object of $M_{star}$ = 2.0 10$^{9}$ M$_\odot$, the typical stellar mass range covered by our sample. These amounts of gas
are re-injected into the ISM in a few Gyr. We can interpolate these values to calculate the mean quantity of recycled gas that should be taken into account for galaxies of different stellar mass. 
With the contribution of the recycled gas, the amount of gas locked in galaxies available to sustain star formation at a rate similar to the present one increases up to  $\tau_{gas,R}$ $\gtrsim$ 3.0-3.3 Gyr. 
These timescales are a factor of $\gtrsim$ 2 longer than the timescale for total gas removal through a ram pressure stripping event. 
They are also comparable to $\sim$ two crossing of the cluster (3.4 Gyr), a time sufficiently long to
allow ram pressure stripping to reach its maximum for $\simeq$ three times. Furthermore, these timescales must
be considered as lower limits to the time necessary to completely stop any star formation activity since the star formation activity smoothly decreases with time after a gradual consumption of the gas content
(Schmidt law). We can thus conclude that, if the timescale for totally removing the gas content of cluster galaxies
is of the order of $\tau_{RP}$ $\simeq$ 1.5 Gyr as indicated by hydrodynamical simulations, ram pressure stripping will be the dominant process regulating
the future activity of these cluster spiral galaxies. \\

Can these results, combined with those obtained in other works, be used to identify the main physical process driving the formation of the bright end of the red sequence in a cluster such as Virgo?  
Ram pressure stripping events have been less efficient in the past because the velocity dispersion within the cluster and the density of the ICM were smaller than at the present epoch. 
It is thus conceivable that their contribution to the quenching of the star formation activity of cluster galaxies has increased with the age of the universe to be at their maximum at the present epoch.
To produce effects on the total gas content comparable to those observed in Virgo H{\sc i}-deficient cluster galaxies ($\mathrm{\hi}-def$ up to 1), the stop of infall of pristine gas from the halo
must have started $\gtrsim$ 7 Gyr ago (Boselli et al. 2006).
This timescale, quite accurate since determined using tuned chemo-spectrophotometric models of galaxy evolution, is even longer than the rough estimates given in this work. 
Here we just indicate that the timescale for gas consumption does not drastically change in different environments. We can thus expect that the timescale for gas consumption via star formation is long at any epoch. 
The same chemo-spectrophotometric models also indicate that 
the effects on the star formation activity of the starved galaxies are minor. They cannot move, for instance, a galaxy from the blue to the red sequence (Boselli et al. 2008a). 
If started at such early epochs, the stop of infall of pristine gas would also have had major effects on the structural properties of the perturbed galaxies. 
Their effective surface brightness would have been much lower than the one of unperturbed, 
spiral galaxies (Boselli et al. 2006). This is exactly the opposite of what is observed (Boselli \& Gavazzi 2006). 
Furthermore, 7 Gyr ago roughly corresponds to the epoch when a cluster such as Virgo started to form
through the aggregation of small groups (Gnedin 2003; McGee et al. 2009; Roediger et al. 2011). Preprocessing was at that time dominating (e.g. Dressler 2004; Dressler et al. 2013). Under these conditions 
the gravitational interactions were much more efficient than those affecting galaxies in a formed cluster because of the low velocity dispersion of the group. 
We thus expect that gravitational interactions were able to dynamically heat the systems, thicken the stellar disc and produce the analogue of local, massive lenticular galaxies.

\section{Conclusion}

We use the recent compilation of homogeneous atomic and molecular gas data for the \textit{Herschel} Reference Survey, a K-band-selected, volume-limited, complete sample of galaxies in the nearby universe,
to study the effects of the surrounding environment on the molecular gas content of cluster galaxies. Using the subsample of isolated, unperturbed objects, 
we first identify the $M(H_2)$ vs. $M_{star}$ relation as the less dispersed scaling relation necessary to estimate the expected molecular gas
content of perturbed galaxies of different size and luminosity. By studying the distribution of the distance to this relation we show that H{\sc i}-deficient, Virgo cluster galaxies
have, on average, less molecular gas than H{\sc i}-normal field objects. 
We then calibrate the H$_2$-deficiency parameter, defined as the difference between the expected and the observed molecular gas mass of an unperturbed galaxy of similar stellar mass.
The H$_2$-deficiency parameter increases with H{\sc i}-deficiency parameter, here taken as tracer of an undergoing perturbation. The relation is weak and scattered, but 
statistically significant (the probability that the two variables are correlated is $P$ $>$ 99 \%). 
This relation is barely detectable in the whole sample because of the large uncertainty on the total CO fluxes due to 
aperture corrections. It is, however, confirmed on a subsample of galaxies with high-quality CO data from mapping. Using a subsample of seventeen CO mapped galaxies, 
we show that, as for the atomic gas, the extension of the molecular disc decreases with increasing H{\sc i}-deficiency. This result suggests that 
gas removal is an outside-in process able to truncate the gaseous disc of galaxies in high-density regions.

The analysis presented in this work shows the presence of molecular gas deficient galaxies in the core of the
Virgo cluster. It is consistent with previous results based on the study of the radial distribution of the molecular gas of a few, resolved galaxies 
(Boselli et al. 2006; Fumagalli et al. 2009) and on the statistical analysis of a heterogeneous sample of cluster and isolated objects (Jablonka et al. 2013). 
This result is statistically significant, based on a homogeneous sample of cluster and field galaxies selected according to similar criteria to minimise selection
biases, and robust vs. the adoption of a constant or a luminosity-dependent conversion factor.
Combined with the recent evidence of dust stripping obtained from \textit{Herschel} observations (Cortese et al. 2010, 2012b), the
deficiency of molecular gas in cluster galaxies can be interpreted as an effect of the ram pressure stripping exerted by the hot intergalactic medium on galaxies moving at high velocity 
within the cluster rather than due to a low formation rate of molecular gas whenever the interstellar medium of galaxies drops below a critical density, as firstly proposed by Fumagalli et al. (2009). 
Given the steep gradient in the molecular gas distribution over the galactic disc, molecular gas stripping is less efficient than atomic gas 
stripping, as indicated by the slope $<$ 1 in the $H_2-def$ vs. $\mathrm{\hi}-def$ relation
and by the steeper slope in the $D(\mathrm{\hi})/D(i)$ vs. $\mathrm{\hi}-def$ (slope = -0.50) than in the $D(CO)/D(i)$ vs. $\mathrm{\hi}-def$ (slope = -0.36) 
relation. The most H{\sc i}-deficient, Virgo cluster galaxies ($\mathrm{\hi}-def$ $\gtrsim$ 1) have, 
on average, a factor of $\sim$ 2 less molecular gas than similar objects in the filed.

The molecular and total gas consumption timescales of these H$_2$-deficient cluster galaxies are $\tau_{H_2}$ $\simeq$ 1.2 Gyr, $\tau_{gas}$ $\simeq$ 2.6 Gyr, and $\tau_{gas,R}$ $\simeq$ 3.0-3.3 Gyr when the recycled gas 
is considered. These timescales are significantly longer than the timescale necessary for ram pressure to produce a typical H{\sc i}-deficient cluster galaxy ($\sim$ 0.1 Gyr). On these timescales
a galaxy can cross the core of the cluster and thus reach the peak of its gas stripping $\simeq$ three times.
The total gas depletion timescale determined considering the contribution of the recycled gas produced by stars after the first crossing of the cluster is also a factor of $\gtrsim$ 2 longer than the typical timescale for 
total gas ablation via ram pressure determined by recent hydrodynamical simulations to be of the order of $\lesssim$ 1.5 Gyr. 
This indicates that ram pressure stripping, rather than gas consumption via star formation, is the main process driving the future evolution of these cluster gas-deficient spiral galaxies. 
The removal of the cold gaseous phase quenches the activity of star formation, inducing the migration of these originally star forming
systems from the blue sequence into the green valley and, eventually into the red sequence.

\begin{acknowledgements}

A.B thanks the ESO visiting program committee for inviting him
at the Garching headquarters for a two-month stay. We are also grateful to M. Fossati
for his help during the preparation of the CO dataset and to M. Gerin for a constructive discussion on the 
timescale for molecular gas formation. We also thank the anonymous referee for useful comments and suggestions 
which improved the quality of the manuscript.
The research leading to these results has received funding from the European Community's Seventh 
Framework Programme (/FP7/2007-2013/) under grant agreement No 229517. 
B.C. is the recipient of an Australian Research Council Future Fellowship (FT120100660).This research has made use of the 
NASA/IPAC Extragalactic Database (NED) 
which is operated by the Jet Propulsion Laboratory, California Institute of 
Technology, under contract with the National Aeronautics and Space Administration
and of the GOLDMine database (http://goldmine.mib.infn.it/).
IRAF is distributed by the National Optical Astronomy Observatory, 
which is operated by the Association of Universities for Research in Astronomy 
(AURA) under cooperative agreement with the National Science Foundation.

\end{acknowledgements}

\begin{table*}
\caption{HRS galaxies with high-quality molecular gas data from Kuno et al. (2007)}
\label{TabKuno}
{
\[
\begin{tabular}{ccccccccccc}
\hline
\noalign{\smallskip}
\hline
HRS	& NGC	& Type			& Member	& $D_{25}$	& log$M_{star}$	& $\mathrm{\hi}-def$	& log$M(H_2)_c$ & log$M(H_2)_v$ & $H_2-def_c$	& $H_2-def$ \\
	&	&			& 		& arcmin	& M$_{\odot}$	&			& M$_{\odot}$	&M$_{\odot}$	&		&		\\
\hline
 36	& 3504	&(R)SAB(s)ab;HII	& Leo Cl.	&  2.69    &  10.24   &    0.60  & 	9.44 &      9.25     & 0.15 & 0.14 \\
 48	& 3631	&SA(s)c			& Ursa Major Cl.&  5.01    &   9.92   &    0.09  &      9.43 &      9.27     &-0.16 &-0.14 \\
 96	& 4212	&SAc;HII		& Virgo N Cl.	&  3.60    &  10.01   &    0.40  &      9.10 &      8.99     & 0.25 & 0.21 \\
102	& 4254  &SA(s)c			& Virgo N Cl.	&  6.15    &  10.39   &    0.06  &     10.02 &      9.73     &-0.28 &-0.23 \\
114	& 4303	&SAB(rs)bc;HII;Sy	& Virgo S Cl.	&  6.59    &  10.51   &    0.05  &      9.94 &      9.63     &-0.08 &-0.03 \\
122	& 4321	&SAB(s)bc;LINER;HII	& Virgo A	&  9.12    &  10.71   &    0.46  &      9.91 &      9.55     & 0.15 & 0.21 \\
149	& 4402	&Sb			& Virgo A	&  3.95    &  10.04   &    0.83  & 	9.31 &      9.23     & 0.07 & 0.00 \\
156	& 4419	&SB(s)a;LINER;HII	& Virgo A	&  3.52    &  10.24   &    1.36  &      9.10 &      8.94     & 0.48 & 0.44 \\
190	& 4501	&SA(rs)b;HII;Sy2	& Virgo A	&  7.23    &  10.98   &    0.58  &      9.88 &      9.50     & 0.45 & 0.47 \\
204	& 4535	&SAB(s)c;HII		& Virgo S Cl.	&  8.33    &  10.45   &    0.32  &      9.55 &      9.33     & 0.24 & 0.22 \\
205	& 4536	&SAB(rs);HII;Sbrst	& Virgo S Cl.	&  7.23    &  10.26   &    0.15  &      9.45 &      9.27     & 0.16 & 0.16 \\
208	& 4548	&SBb(rs);LINER;Sy	& Virgo A  	&  6.00    &  10.74   &    0.94  &      8.88 &      8.61     & 1.21 & 1.18 \\
217	& 4569	&SAB(rs)ab;LINER;Sy	& Virgo A	& 10.73    &  10.66   &    1.05  &      9.68 &      9.35     & 0.32 & 0.37 \\
220	& 4579	&SAB(rs)b;LINER;Sy1.9	& Virgo A	&  6.29    &  10.94   &    0.98  &      9.36 &      9.01     & 0.93 & 0.94 \\
247	& 4654	&SAB(rs)cd;HII		& Virgo E Cl.	&  4.99    &  10.14   &   -0.05  &      9.61 &      9.43     &-0.13 &-0.12 \\
254	& 4689	&SA(rs)bc		& Virgo E Cl.	&  5.86    &  10.19   &    0.99  & 	9.31 &      9.22     & 0.22 & 0.13 \\
295	& 5248	&(R)SB(rs)bc;Sy2;HII	& Virgo-Lybra Cl.& 1.79    &  10.43   &    0.08  & 	9.77 &      9.61     & 0.00 &-0.07 \\
\noalign{\smallskip}
\hline
\end{tabular}
\]
}
\end{table*}

\begin{table*}
\caption{Calibration of the H$_2$-deficiency parameter: coefficients of the molecular gas-variable relations: log$M(H_2)$ = $c$ $\times$ Variable + $d$ (Figs. \ref{calraggi}, \ref{calmass}, and \ref{calkuno})}
\label{Tabcaldeffit}
{
\[
\begin{tabular}{cccccccc}
\hline
\hline
\noalign{\smallskip}
$X_{CO}$	& Variable			&Sample$^1$	& c		& d 	& $\rho^2$	& $\sigma^3$	& N.obj.\\
\hline
constant	& log[$\pi r_{24.5}(g)^2$]	&all		& 0.75$\pm$0.11 & 7.02$\pm$0.26 & 0.51  & 0.34		& 96 \\
lum.dep.	&				&all		& 0.43$\pm$0.10 & 7.74$\pm$0.23 & 0.37  & 0.35		& 96 \\
\hline
constant	& log$M_{star}$			&all		& 0.81$\pm$0.07	& 0.84$\pm$0.73	& 0.69	& 0.27		& 95 \\ 
lum.dep.	& 				&all		& 0.49$\pm$0.07	& 3.95$\pm$0.71	& 0.52  & 0.31		& 95 \\ 
\hline
constant	& log$M_{star}$			&Kuno		& 1.01$\pm$0.37	&-0.76$\pm$3.79	& 0.69	& 0.14		&  8 \\ 
lum.dep.	& 				&Kuno		& 0.80$\pm$0.32	& 1.24$\pm$3.29	& 0.62  & 0.13		&  8 \\ 
\noalign{\smallskip}
\hline
\end{tabular}
\]
Notes: linear fit.\\
$1$: The sample used to fit the $M(H_2)$ vs. $Variable$ relations is composed of all spiral galaxies
($all$) with molecular gas (considering CO non-detections as detections) with and H{\sc i} 
deficiency parameter $\mathrm{\hi}-def$ $\leq$ 0.4 (black filled dots and triangles in Figs. \ref{calraggi} and \ref{calmass}). The sample $Kuno$ indicates the subsample of 
galaxies with integarted CO data from Kuno et al. (2007). For this subsample, the galaxies used to fit the 
relation are all those with $\mathrm{\hi}-def$ $\leq$ 0.4 (red filled dots in Fig. \ref{calkuno}).\\
$2$: Spearman correlation coefficient.\\
$3$: Dispersion in the relation.\\
}
\end{table*}

\begin{table*}
\caption{Mean values and standard deviations of the molecular gas-optical surface and stellar mass relations (Figs. \ref{calraggi}, \ref{calmass}, and \ref{calkuno})}
\label{Tabcaldefdata}
{
\[
\begin{tabular}{ccccccc}
\hline
\hline
\noalign{\smallskip}
\multicolumn{4}{c}{}&\multicolumn{1}{c}{$X_{CO}$ constant}&\multicolumn{1}{c}{$X_{CO}$ lum.dep.}&\multicolumn{1}{c}{}\\
x			&	y	&  Sample	& $<x>$ &	 $<y>$   & $<y>$	 & N  \\
\hline
log[$\pi r_{24.5}(g)^2$]& log$M(H_2)$	& all		& 1.79$\pm$0.19	 & 8.48$\pm$0.25 & 8.60$\pm$0.25 & 20 \\
			&		& 		& 2.26$\pm$0.13	 & 8.59$\pm$0.44 & 8.61$\pm$0.38 & 47 \\
			&		& 		& 2.69$\pm$0.13	 & 9.15$\pm$0.53 & 8.98$\pm$0.46 & 24 \\
			&		& 		& 3.26$\pm$0.16	 & 9.51$\pm$0.17 & 9.19$\pm$0.09 & 5 \\
\noalign{\smallskip}
\hline
log$M_{star}$	& log$M(H_2)$		& all		& 9.15$\pm$0.10	 & 8.37$\pm$0.39 & 8.53$\pm$0.41 & 16 \\
		&			&  		& 9.48$\pm$0.13	 & 8.45$\pm$0.21 & 8.53$\pm$0.18 & 34 \\
		&			& 		& 9.99$\pm$0.14	 & 8.93$\pm$0.39 & 8.82$\pm$0.37 & 30 \\
		&			& 		&10.51$\pm$0.12	 & 9.48$\pm$0.50 & 9.21$\pm$0.48 & 13 \\
\hline
log$M_{star}$	& log$M(H_2)$		& Kuno		& 10.02$\pm$0.11 & 9.38$\pm$0.26 & 9.23$\pm$0.22 & 3 \\
		&			& 		& 10.41$\pm$0.09 & 9.75$\pm$0.24 & 9.51$\pm$0.21 & 5 \\
\noalign{\smallskip}
\hline
\end{tabular}
\]
Note: scaling relations for unperturbed ($\mathrm{\hi}-def$ $\leq$ 0.4) spiral galaxies determined considering H{\sc i} and CO non-detections as detections.
}
\end{table*}

\begin{table*}
\caption{Coefficients of the H{\sc i}-H$_2$ deficiency relation: $H_2-def$=$a$ $\times$ $\mathrm{\hi}-def$ + $b$ (Fig. \ref{defHIdefH2kuno})}
\label{Tabdefdeffit}
{
\[
\begin{tabular}{cccccccccc}
\hline
\noalign{\smallskip}
\hline
$X_{CO}$	& Sample				& Calibration 	& a	& b 	& $\rho^1$ & $\sigma$ $^2$ & Probability$^3$ & Symbol & Fig. \\
\hline
constant	& Kuno	    				& Kuno		& 0.88  & -0.22 & 0.77	& 0.22	& $>$99.9\%	& red squares	& \ref{defHIdefH2kuno}c  \\ %
lum.dep.	& Kuno					& Kuno		& 0.87	& -0.23	& 0.71  & 0.23	& $>$99.9\%	& red squares 	& \ref{defHIdefH2kuno}d  \\ %
\hline
\noalign{\smallskip}
\hline
\end{tabular}
\]
Notes: bisector fit considering CO and H{\sc i} non-detections at their upper limit.\\
$1$: Spearman correlation coefficient.\\ 
$2$: Dispersion in the relation.\\
$3$: Probability that the two variables are correlated.\\
}
\end{table*}

\begin{table*}
\caption{Mean values and standard deviations of the H$_2$ vs. H{\sc i} deficiency parameter relations (Figs. \ref{defHIdefH2} and \ref{defHIdefH2kuno})}
\label{Tabdefdefdata}
{
\[
\begin{tabular}{cccccccc}
\hline
\hline
\noalign{\smallskip}
$X_{CO}$	& Sample	& Calibration 	& Symbol	& Fig. 			& $<\mathrm{\hi}-def>$	& $<H_2-def>$	& N	\\
\hline
constant	& all		& all		& blue		& \ref{defHIdefH2}a	& -0.01$\pm$0.20	& 0.00$\pm$0.36	& 80 \\
		&		&		&		&			& 0.43$\pm$0.09		& 0.04$\pm$0.37 & 34 \\
		&  		&		&		&			& 0.73$\pm$0.07		& 0.17$\pm$0.36 & 16 \\
		&		&		&		&			& 1.03$\pm$0.10		& 0.29$\pm$0.38 & 18 \\
		&		&		&		&			& 1.42$\pm$0.10		& 0.36$\pm$0.29 & 13 \\
\hline
lum.dep.	& all		& all		& blue		& \ref{defHIdefH2}b	& -0.01$\pm$0.20	& 0.00$\pm$0.34	& 80 \\
		&		&		&		&			& 0.43$\pm$0.09		& 0.01$\pm$0.37 & 34 \\
		&  		&		&		&			& 0.73$\pm$0.07		& 0.11$\pm$0.35 & 16 \\
		&		&		&		&			& 1.03$\pm$0.10		& 0.23$\pm$0.39 & 18 \\
		&		&		&		&			& 1.42$\pm$0.10		& 0.27$\pm$0.28 & 13 \\
\hline
constant	&$M_{star}>10^{10}$ M$_{\odot}$ & all		& blue		& \ref{defHIdefH2}a		& 0.07$\pm$0.16		&-0.03$\pm$0.46	& 21 \\
		&		&		&		&						& 0.45$\pm$0.09		&-0.04$\pm$0.33 & 13 \\
		&  		&		&		&						& 0.74$\pm$0.09		& 0.23$\pm$0.37 &  5 \\
		&		&		&		&						& 1.03$\pm$0.09		& 0.38$\pm$0.36 & 13 \\
		&		&		&		&						& 1.39$\pm$0.08		& 0.32$\pm$0.25 &  8 \\
\hline
lum.dep.	&$M_{star}>10^{10}$ M$_{\odot}$ & all		& blue		& \ref{defHIdefH2}b		& 0.07$\pm$0.16		&-0.02$\pm$0.44	& 21 \\
		&		&		&		&						& 0.45$\pm$0.09		&-0.07$\pm$0.30 & 13 \\
		&  		&		&		&						& 0.74$\pm$0.09		& 0.15$\pm$0.36 &  5 \\
		&		&		&		&						& 1.03$\pm$0.09		& 0.33$\pm$0.36 & 13 \\
		&		&		&		&						& 1.39$\pm$0.08		& 0.41$\pm$0.26 &  8 \\
\hline
constant	& Kuno		& Kuno		& cyan		& \ref{defHIdefH2kuno}c	& 0.06$\pm$0.07		& -0.08$\pm$0.15&  6 \\
		&		&		&		&			& 0.47$\pm$0.12		& 0.25$\pm$0.12 &  5 \\
		&  		&		&		&			& 0.99$\pm$0.05		& 0.67$\pm$0.47 &  4 \\
\hline
lum.dep.	& Kuno		& Kuno		& cyan		& \ref{defHIdefH2kuno}d	& 0.06$\pm$0.07		& -0.07$\pm$0.13&  6 \\
		&		&		&		&			& 0.47$\pm$0.12		& 0.25$\pm$0.13 &  5 \\
		&  		&		&		&			& 0.99$\pm$0.05		& 0.65$\pm$0.49 &  4 \\ 
\hline
\noalign{\smallskip}
\hline
\end{tabular}
\]
Note: mean values determined considering H{\sc i} and CO non-detections at their upper limit.}
\end{table*}

\begin{table*}
\caption{Mean gas depletion timescales and standard deviations for H{\sc i}-normal ($\mathrm{\hi}-def$ $\leq$ 0.4) and H{\sc i}-deficient ($\mathrm{\hi}-def$ $>$ 0.4) galaxies.}
\label{Tabtau}
{
\[
\begin{tabular}{cccc}
\hline
\noalign{\smallskip}
\hline
			&		& Sample			& Sample \\
Dep. Time		& $X_{CO}$	& $\mathrm{\hi}-def$ $\leq$ 0.4	& $\mathrm{\hi}-def$ $>$ 0.4 \\
			&		& yr				&	yr	\\
\hline
log($\tau_{H_2}$)	&constant	&  8.81$\pm$0.36		& 9.12$\pm$0.39 \\ 
  	  		&lum.dep.	&  8.78$\pm$0.33		& 9.05$\pm$0.39 \\ 
log($\tau_{gas}$)  	&constant	&  9.61$\pm$0.24 		& 9.44$\pm$0.29 \\ 
  	  		&lum.dep.	&  9.59$\pm$0.26		& 9.39$\pm$0.32 \\ 
log($\tau_{gas,R}$)	&constant	&  9.63$\pm$0.24		& 9.52$\pm$0.31 \\ 
  	  		&lum.dep.	&  9.61$\pm$0.26		& 9.48$\pm$0.32 \\ 			
\noalign{\smallskip}
\hline
\end{tabular}
\]
}
\end{table*}

\end{document}